\documentclass[12 pt]{article}
\counterwithin*{equation}{section}
\usepackage[all]{xy}
\usepackage{amsmath,amssymb,amsfonts,latexsym,float,graphics,epsfig}
\usepackage{accents}

\usepackage{setspace}
\usepackage{xcolor}
\usepackage[top=2cm, bottom=2cm, left=2cm, right=2cm]{geometry}
\usepackage[colorlinks]{hyperref}
\usepackage{tikz}
\usetikzlibrary{cd}

\renewcommand\theequation{\arabic{section}.\arabic{equation}}

\newcommand{\p}{\partial}

\begin{document}
\title{Conformal and Contact Kinetic Dynamics and Their Geometrization} \maketitle

\begin{center}

\author{O\u{g}ul Esen\footnote{oesen@gtu.edu.tr, Department of Mathematics, Gebze Technical University,   41400 Gebze-Kocaeli, Turkey \\ 
Center for Mathematics and its Applications,
Khazar University, Baku, AZ1096, Azerbaijan},   Ayten Gezici\footnote{*Corresponding author, agezici@gtu.edu.tr,  Department of Mathematics, Gebze Technical University,  \\  41400 Gebze-Kocaeli, Turkey}*,  
Miroslav Grmela\footnote{miroslav.grmela@polymtl.ca, \'{E}cole Polytechnique de Montr\'{e}al,
  C.P.6079 suc. Centre-ville, 
 Montr\'{e}al, H3C 3A7,  Qu\'{e}bec, Canada}, 
Hasan Gümral\footnote{hgumral@yeditepe.edu.tr, Department of Mathematics, Yeditepe University,  34755 Ataşehir-İstanbul, Turkey},
 Michal Pavelka\footnote{pavelka@karlin.mff.cuni.cz, Mathematical Institute, Faculty of Mathematics and Physics, Charles University, 
 Sokolovsk\'{a} 83, 18675 Prague, Czech Republic}  \\   
and Serkan Sütlü\footnote{serkansutlu@gtu.edu.tr, Department of Mathematics, Gebze Technical University,   41400 Gebze-Kocaeli, Turkey}
}
\end{center}


\bigskip

\begin{abstract}
We propose a conformal generalization of the reversible Vlasov equation of
kinetic plasma dynamics, called conformal kinetic theory. In order to arrive at
this formalism, we start with the conformal Hamiltonian dynamics of particles
and lift it to the dynamical formulation of the associated kinetic theory. The
resulting theory represents a simple example of a geometric pathway from
dissipative particle motion to dissipative kinetic motion. We also derive the
kinetic equations of a continuum of particles governed by the contact
Hamiltonian dynamics, which may be interpreted in the context of relativistic
mechanics. Once again we start with the contact Hamiltonian dynamics and lift
it to a kinetic theory, called contact kinetic dynamics. Finally, we project
the contact kinetic theory to conformal kinetic theory so that they form a
geometric hierarchy. 

\smallskip

\smallskip


\end{abstract}

\tableofcontents
\setlength{\parskip}{4mm}

\onehalfspacing

\section{Introduction}

The dynamics of a non-relativistic and collisionless plasma resting in $Q\subset \mathbb{R}^3$ is determined by the plasma density function $f=f(q^i,p_i)$, defined on the momentum phase space $T^*Q$ with Darboux' coordinates $(q^i,p_i)$. Equation of motion then is a coupled integrodifferential system 
\begin{equation}\label{PV}
\begin{split}
\frac{\partial f}{\partial t}+\frac{1}{m}p_{i}\frac{\partial f}{\partial
q^{i}}-e\frac{\partial \phi }{\partial q^{i}}\frac{\partial f}{\partial
p_{i}}=0\\ 
\nabla _{q}^{2}\phi _{f}(q)=-e\int f(q,p)d^{3}p
\end{split}
\end{equation}
which are known as the Vlasov–Poisson equations, where $e$ is the charge and $\phi$ is the potential. Hamiltonian analysis of this system may be recalled from \cite{marsden1982hamiltonian,morrison1981hamiltonian}, wherein it is well established that the Vlasov–Poisson system \eqref{PV} admits Hamiltonian formulation. More precisely,  the Vlasov equation fits in Lie-Poisson (a Poisson framework available on the dual of a Lie algebra \cite{Marsden1999}) picture. 

In a series of papers \cite{EsGu12,Gu10}, while investigating Lie-Poisson formulation of the Vlasov equation, an intermediate level of description  is obtained on the space of one-forms on $T^*Q$. In this case,  the dynamics is represented by the evolution of a dual element, more precisely a one-form $\Pi$, governed by Hamiltonian vector field $X_H$ through 
\begin{equation}\label{MV-intro}
\dot{\Pi}=-\mathcal{L}_{X_H}\Pi
\end{equation}
where $\mathcal{L}_{X_H}$ is the Lie derivative, whereas $H=p^2/2m+e\phi$ is assumed to be the total energy of a single particle. The link between the Vlasov equation (the first line in  \eqref{PV}) and the momentum-Vlasov equation \eqref{MV-intro} is determined through the dual mapping of Lie algebra homomorphism $H\mapsto X_H$ which is computed to be 
\begin{equation}\label{density}
f=\mathrm{div}~\Omega^\sharp_Q(\Pi).
 \end{equation}
Here, $\Omega_Q$ is the canonical symplectic two-from on $T^*Q$, and $\Omega^\sharp_Q$ is the induced musical isomorphism, while $\mathrm{div}$ denotes the symplectic divergence. Since \eqref{density} is determined as the dual of a Lie algebra homomorphism, it is both a Poisson and a momentum map. In this geometrization,  the Poisson equation (the second line  in  \eqref{PV}) is realized as the momentum map due to the gauge invariance of Hamiltonian dynamics. One of the interesting features of the momentum-Vlasov equation \eqref{MV-intro} is its pure geometric derivation. More precisely, starting from a Hamiltonian vector field   $X_H$ and then lifting it to the cotangent bundle and later  taking to the vertical representative, one arrives at a generalized vector field $V\widehat{X}_H$ which determines the motion of $\Pi$ as given in \eqref{MV-intro}. Keeping the same line of thought,  further analysis has also been carried out on fluid dynamics \cite{EsGrMiGu19,EsGu11}. The explicit time dependent analysis of the Vlasov dynamics in a geometric formalism is given in \cite{atecsli2023non}. Additionally, rich algebraic structure of momentum-Vlasov dynamics is examined in \cite{EsSu21}, inspired from the moment algebra of Vlasov dynamics, see for instance \cite{gibbons1981collisionless,gibbons2008vlasov,holm2009geodesic}. We cite a recent work \cite{esen2024variational}, which studies the locally conformal character of Vlasov dynamics, similar to the approach in this work. That study considers the continuous locally conformal geometric setting, based on the discrete picture from \cite{MR4747740}.

This present work consists of two main sections and a conclusion section, in which we propose novel geometries and kinetic theories generalizing the ones in the literature, along with an appendix.

\textbf{Section \ref{Sec-Sym-Man}: Conformal Kinetic Dynamics.} Classical Hamiltonian vector fields on symplectic manifolds are divergence-free. This is one of the manifestations of reversibility. In \cite{mclachlan2001conformal}, Hamiltonian vector fields are generalized to conformal vector fields with constant divergences. 
In this work, our first goal is to present a kinetic equation of particles governed by conformal vector fields. This novel generalization will be done both on the dynamics of one-form section as well as the dynamics of density function (the link between these two realizations, on the other hand, will be established by a Poisson map in Appendix \ref{link-link}). We shall also provide a geometrical pathway from the particle motion to the motion of the continuum by means of geometrical operations such as complete lifts and vertical representatives. We note that, in this work, we are interested in the conformal Hamiltonian dynamics defined on the cotangent bundle which is the generic picture (see \cite{McLachlan2001})  of conformal framework. This choice enables us to employ Liouville vector field so that more explicit calculations.

\textbf{Section \ref{Sec:kin-con}: Contact Kinetic Dynamics.} Even though contact manifolds are known as the odd-dimensional counterparts of symplectic manifolds, there exist some characteristic differences. In the contact framework, Hamiltonian flow preserves neither the Hamiltonian function nor the volume form.  Our second goal is to provide the kinetic dynamics of a bunch of particles under contact Hamiltonian motion. So, dissipative motion on the particle level gives rise to dissipative motion on the level of density functions. 

\textbf{Conclusion Section \ref{Sec:Hie}: A Hierarchy from Contact to Conformal
Dynamics.} To sum up the discussions, we shall present the hierarchy of the underlying Lie algebras of the previous sections. We shall later dualize the Lie algebra homomorphisms to arrive at momentum and Poisson mappings connecting kinetic dynamics on different levels of descriptions, namely, the reversible Hamiltonian dynamics, conformal Hamiltonian dynamics, and the contact Hamiltonian dynamics.  

\textbf{Notation.} We consider the manifold where the plasma particles are at rest as $ Q \subset \mathbb{R}^3 $, with coordinates $ (q^i) $. The cotangent bundle, which is the canonical symplectic manifold, $ T^*Q $, is a 6-dimensional manifold with Darboux coordinates $ (q^i, p_i) $, representing the momentum phase space.

In the manuscript, we refer to an arbitrary manifold, which in most cases is assumed to be symplectic, by $ M $. In this case, we denote the local coordinates by $ (x^a) $. The cotangent bundle of $ M $ is $ T^*M $, and the Darboux coordinates are written as $ (x^a, y_a) $. Specifically, if $ M = T^*Q $, then the coordinates on $ M $ can be written in Darboux form as $ (x^a) \mapsto (q^i, p_i) $. Furthermore, the Darboux coordinates on $ T^*M = T^*T^*Q $ are written as $ (x^a, y_a) \mapsto (q^i, p_i, \Pi_i, \Pi^i) $.

For the tensorial objects over the manifolds, we shall follow the notation used in the literature, see \cite{abraham1978foundations,leon89,holm11,holm2009geometric, LiMa87,
Marsden1999}. More precisely, given a manifold $M$, we shall denote the space of smooth functions by $\mathcal{F}(M)$, the space of one-form sections by $\Lambda^1(M)$, and the space of vector fields by $\mathfrak{X}(M)$. 
On the other hand, given a two-form $\Omega$, we shall make use of the musical flat mapping 
\begin{equation} \label{flat}
\Omega^\flat: \mathfrak{X}(M)\longrightarrow \Lambda^1(M), \qquad \Omega^\flat(X) (Y)=(\iota_X \Omega) (Y): = \Omega(X,Y).
\end{equation}
Moreover, in case $\Omega^\flat$ is invertible (occurs if $\Omega$ is non-degenerate), we shall denote its inverse by $\Omega^\sharp$.  Finally, given a one-form $\alpha$ and a vector field $X$, we shall represent by
\begin{equation}\label{Omega-sharp}
\alpha(X)=\Omega(\Omega^{\sharp }(\alpha),X)
\end{equation}
the relation between the symplectic two-form $\Omega$, and the musical mapping $\Omega^{\sharp }$.

\section{Conformal Kinetic Theory}\label{Sec-Sym-Man}

 In this section, we shall consider a particle that is governed by a conformal vector field that dissipates the energy and has non-zero divergence. We note that in this section, the dissipation arises from a conformal factor, which is a real constant, not a function of the state variables. We then lift this particle  motion to a kinetic theory underlying the dynamics of a number of such particles.
 
\subsection{Conformal Hamiltonian Dynamics}

Let $M$ be a manifold. It is called a symplectic manifold if it admits a closed non-degenerate two form $\Omega$, \cite{Mars88,Weinstein-symplectic}. Accordingly, we shall denote a symplectic manifold by a pair $(M,\Omega)$.  The non-degeneracy of $\Omega$ suffices to define a non-vanishing top form on the manifold, called the symplectic volume 
 \begin{equation}\label{symp-volume}
d \mu=\frac{(-1)^{n(n-1)/2}}{n!} \Omega\wedge \dots\wedge \Omega.
 \end{equation} 
 The generic example of a symplectic manifold is the cotangent bundle $T^*Q$ of a manifold $Q$ (in this work, $Q\subset \mathbb{R}^3$), along with the Liouville one form $\Theta_Q$ and the canonical symplectic two-form $\Omega_Q=-d\Theta_Q$.

\textbf{Classical Hamiltonian Dynamics.} Given a Hamiltonian function $H$, the Hamiltonian vector field $X_{H}$ is defined through the Hamilton's equation 
\begin{equation}
\iota_{X_{H}}\Omega =dH.  \label{Hamvf}
\end{equation}
Taking the exterior derivative of (both sides of) the equality, we see that the Lie derivative of the symplectic two-form vanishes. As such, the divergence of a Hamiltonian vector field is zero concerning the symplectic volume \eqref{symp-volume}. We record these as 
\begin{equation}
\mathcal{L}_{X_{H}}\Omega =0, \qquad  \mathcal{L}_{X_{H}} d \mu =0 \label{Hamvf-Lie},
\end{equation}
respectively. Next, the integration of \eqref{Hamvf-Lie} implies that the integral flows $\varphi_t$ of the Hamiltonian vector field preserve both the symplectic two-form and the symplectic volume 
\begin{equation}
\varphi_t^*(\Omega)=\Omega,\qquad \varphi_t^*( d \mu)= d \mu, 
\end{equation}
respectively. Furthermore, the calculations 
\begin{equation}
\mathcal{L}_{X_{H}}(H)=X_{H}(H)=0, \qquad \varphi_t^*(H)=H
\end{equation}
show us that the Hamiltonian function is constant along the motion. This corresponds to the conservation of energy and determines the reversible character of the symplectic Hamiltonian dynamics. 

A symplectic two-form can be used to determine a  Poisson bracket on the space of smooth functions given by 
\begin{equation}\label{poissonsymplectic}
\{F,H\}^{(S)}:= \Omega\big(X_{F},X_{H}\big),
\end{equation} 
which is skew-symmetric and satisfies both the Leibniz and the Jacobi identities. Since the characteristic distribution is integrable, the space of Hamiltonian vector fields is closed under the Jacobi-Lie bracket. More precisely we have
\begin{equation}\label{H-XH}
[X_F,X_H]=-X_{\{F,H\}^{(S)}}.
\end{equation}
In other words,
\begin{equation}
 \mathfrak{X}_{\mathrm{ham}}(M):=\{X_H\in \mathfrak{X}(M):  \iota_{X_{H}}\Omega =dH \}
\end{equation} 
is a Lie algebra.

\textbf{Conformal Hamiltonian Dynamics.} 
A Hamiltonian vector field on $M$ is defined through the covariance equation \eqref{Hamvf}. Recall from \eqref{Hamvf-Lie} that the Lie derivative vanishes identically and the divergence is zero. Relaxing this condition, we define a conformal vector field $X_H^c$ as \cite{Carinena13,mclachlan2001conformal,McLachlan2001}
\begin{equation}\label{conf-Ham-1}
\mathcal{L}_{X_H^c}\Omega=c \Omega
\end{equation}
for a fixed real number $c$  called  conformal parameter. In the sequel, we shall work on several conformal vector fields making use of subindexes (such as $c_H$) to be more precise about the parameter. Let us note also that a Hamiltonian vector field is a conformal vector field with the conformal factor being zero. The divergence of a conformal vector field is non-zero. More precisely (for a symplectic manifold of dimension $2n$) it is computed to be 
\begin{equation}\label{div-X-c}
\mathcal{L}_{X_H^c} d \mu= cn  d \mu,\qquad \mathrm{div}(X_H^c)=nc,
\end{equation}
where $d \mu$ is the symplectic volume. Integration of the defining identity \eqref{conf-Ham-1} yields that the flow of a conformal  vector field preserves the symplectic two-form up to the conformal factor $c$ and the volume up to  the conformal factor $cn$ that are
\begin{equation}
\varphi_t^*(\Omega)=\exp(ct)\Omega,\qquad \varphi_t^*( d \mu)=\exp(nct)d \mu,
\end{equation}
respectively, where $\exp$ stands for the exponential function. 

For an exact symplectic manifold where $\Omega=-d\Theta$,  we define a conformal  vector field as
\begin{equation}\label{conf-Ham-2}
\iota_{X_H^c}\Omega=dH-c\Theta.
\end{equation}
For the exact symplectic manifolds, as announced in \cite{McLachlan2001}, the formulation \eqref{conf-Ham-2} is generic for conformal vector fields. That is the conformal vector field $X_H^c$ defined in \eqref{conf-Ham-1} admits a function $H$ realizing the covariance equation \eqref{conf-Ham-2}. Further, again by citing \cite{McLachlan2001}, if the first cohomology of the manifold is vanishing then $X_H^c$ is written as a combination of the symplectic Hamiltonian vector field $X_H$ and the Liouville vector field $Z$. In this work, we consider the conformal Hamiltonian dynamics defined on the cotangent bundle which is exact symplectic.  

Following \cite{LiMa87}, we define the Liouville vector field $Z$ as the image of the Liouville (potential) one-form $\Theta$ under the musical isomorphism $\Omega^\sharp$ (recall from \eqref{flat} and \eqref{Omega-sharp}) induced from the symplectic form $\Omega$, namely,
\begin{equation}\label{Liouville-field}
Z:=\Omega^\sharp(\Theta),\qquad \Theta=\Omega^\flat(Z).
\end{equation}
It is worth noting that the Liouville vector field is not a Hamiltonian vector field but a conformal vector field with the conformal factor $-1$, and with the divergence $-n$, as 
\begin{equation}\label{divergence-Z}
\mathcal{L}_Z\Omega=-\Omega, \qquad \mathcal{L}_Zd \mu=-nd \mu, \qquad \mathrm{div}(Z)=-n.
\end{equation}
In terms of the Liouville vector field, defined in \eqref{Liouville-field}, we can express a conformal vector field as the linear combination
\begin{equation}\label{c-reel}
X_H^c=X_H-cZ,
\end{equation}
where $X_H$ is the Hamiltonian vector field for the Hamiltonian function. Accordingly, we compute the change of the Hamiltonian function along the conformal  vector field, and its flow $\phi_t$, as
\begin{equation}
\mathcal{L}_{X_{H}^c}(H)=X_{H}(H)-cZ(H)=-cZ(H), \qquad \phi_t^*(H)=H-c\psi_t^*(H),
\end{equation}
respectively, where $\psi_t$ denotes the integral curve of the Liouville vector field $Z$. 

We consider the Darboux' coordinates $(q^i,p_i)$ on $M$ assuming that it is locally isomorphic to the momentum phase space $M=T^*Q$ of the configuration manifold $Q$. The symplectic Hamiltonian vector field $X_H$ and the Liouville vector field $Z$ are computed to be
   \begin{equation}
   X_H= \frac{\partial H}{\partial p_i}\frac{\partial}{\partial q^i} - \frac{\partial H}{\partial q^i}\frac{\partial}{\partial p_i},\qquad 
Z=-p_i\frac{\partial}{\partial p_i},
           \end{equation} 
           respectively. So the conformal vector field $X_H^c$ becomes 
   \begin{equation}
   X_H^c= \frac{\partial H}{\partial p_i}\frac{\partial}{\partial q^i} - \big( \frac{\partial H}{\partial q^i}-cp_i\big) \frac{\partial}{\partial p_i}.
           \end{equation} 
Then, the dynamics governed by a conformal vector field $X_H^c$ is given by
        \begin{equation} \label{conformal-Ham-eq}
        \dot{q}^i=\frac{\partial H}{\partial p_i},\qquad \dot{p}_i=-\frac{\partial H}{\partial q^i}+cp_i.
        \end{equation} 
Let us note that in case the conformal factor $c$ is zero,  \eqref{conformal-Ham-eq} reduces to the classical Hamilton's equations as expected. 

\subsection{Lie Algebra of Conformal Vector Fields} 

Consider a symplectic manifold $(M,\Omega)$. 
Let us first note that the Jacobi-Lie bracket of two conformal   vector fields  is a local Hamiltonian vector field. Indeed,
\begin{equation}\label{X-c-hom}
\mathcal{L}_{[X_H^c,X_K^c]}\Omega=\mathcal{L}_{X_H^c}\mathcal{L}_{X_K^c}\Omega - \mathcal{L}_{X_K^c}\mathcal{L}_{X_H^c} \Omega=(c_H c_K - c_K c_H) \Omega =0.
\end{equation}
As such, we can argue that the space of conformal  vector fields is a Lie algebra 
\begin{equation}
\mathfrak{X}_{\mathrm{ham}}^c(M):=\{X_H^c\in \mathfrak{X}(M):  \mathcal{L}_{X_{H}^c}\Omega =c_H \Omega \}
\end{equation}
and contains the space of Hamiltonian vector fields $\mathfrak{X}_{\mathrm{ham}}(M)$ as an ideal.

Let, now, $\mathfrak{z}$ denotes the space of vector fields spanned by scalar multiples (that is $\mathbb{R}$) of the Liouville vector field $Z$. Evidently, this space is one-dimensional over $\mathbb{R}$ and may be obtained as
\begin{equation}
\mathbb{R}\longleftrightarrow \mathfrak{z},\qquad c\leftrightarrow cZ,
\end{equation}
and thus acquires the structure of a trivial Lie algebra. More precisely, being the linear span (with $\mathbb{R}$) of $Z$, $\mathfrak{z}$ is a trivial Lie subalgebra of the one-dimensional projective module over the ring of functions generated by the vector field $Z$. 
Moreover, $\mathfrak{z}$ acts on the space of Hamiltonian vector fields $\mathfrak{X}_{\mathrm{ham}}(M)$ from the left as  
\begin{equation} \label{Z-action}
\mathfrak{z} \times \mathfrak{X}_{\mathrm{ham}}(M)  \longrightarrow \mathfrak{X}_{\mathrm{ham}}(M),\qquad (Z,X_H)\mapsto [Z,X_H]=X_{Z(H)+H}.
\end{equation}

The realization \eqref{c-reel} motivates us to have the space $\mathfrak{X}_{\mathrm{ham}}^c(M)$ of conformal   vector fields as the Cartesian product of the space $\mathfrak{X}_{\mathrm{ham}}(M)$ of Hamiltonian vector fields and $\mathfrak{z}\simeq \mathbb{R}$. 
Accordingly, we can recast the space $\mathfrak{X}_{\mathrm{ham}}^c(M)$ of conformal vector fields as a central extension of the space $\mathfrak{X}_{\mathrm{ham}}(M)$ of Hamiltonian vector fields as
\begin{equation}\label{X_Ham-c}
\mathfrak{X}_{\mathrm{ham}}(M) \times  \mathfrak{z}
\longleftrightarrow \mathfrak{X}_{\mathrm{ham}}^c(M) 
 ,\qquad(X_H,c)\mapsto X_H^c=X_H-cZ .
\end{equation}

Now we are ready to determine the Lie algebra structure on $\mathfrak{X}^c_{\rm{ham}}(M)$. To this end, given two conformal fields $X_H^c$ and $X_F^c$, we compute their Lie bracket
\begin{equation}\label{conf-Lie-hom}
\begin{split}
[X_H^c,X_F^c]_{\mathfrak{X}}=-
[X_H^c,X_F^c]&=-[X_H-c_HZ,X_F-c_FZ]
\\&=-[X_H,X_F]+c_H[Z,X_F]+c_F[X_H,Z]
\\&=X_{\{H,F\}^{(S)}}+c_HX_{Z(F)+F}-c_FX_{Z(H)+H}
\\&=X_{\{H,F\}^{(S)}+c_H(Z(F)+F)-c_F(Z(H)+H)},
\end{split}
\end{equation}
where we have employed the action \eqref{Z-action} on the fourth equality. 

Let us conclude the present subsection with another characterization of conformal vector fields that will be useful in the sequel. 

To this end, let us recall that we have identified the space $\mathfrak{X}_{\mathrm{ham}}(M) $ of Hamiltonian vector fields with the space $\mathcal{F}(M)$ of smooth functions (modulo constants). We now employ this to the identification in \eqref{X_Ham-c} to obtain 
\begin{equation}\label{Phi-c}
 \Phi ^c: \mathcal{F}(M) \times \mathfrak{z} \longrightarrow\mathfrak{X}_{\mathrm{ham}}^c(M) 
 ,\qquad (H, c)\mapsto X_H^c=X_H-cZ.
\end{equation}
In view of \eqref{conf-Lie-hom}, it is thus possible to endow $\mathcal{F}(M) \times \mathfrak{z} $ with a Lie algebra structure so that the mapping $\Phi ^c$ is a Lie algebra homomorphism. Accordingly, we define the bracket 
 \begin{equation}\label{alg-ext}
 [(H,c_H),(F,c_F)]:=\big( \{H,F\}^{(S)}+c_H(Z(F)+F)-c_F(Z(H)+H),0\big)
  \end{equation}
which happens to be a Lie algebra bracket, satisfying the Jacobi identity.

 \subsection{Conformal Kinetic Dynamics in Momentum Formulation} 

In order to characterize the dual space $\mathfrak{X}_{\mathrm{ham}}^{c*}(M)$ of the space $ \mathfrak{X}_{\mathrm{ham}}^c(M)$ of conformal vector fields we shall now consider the Lie algebra $ \mathcal{F}(M) \times \mathfrak{z} $. 

For the function space $\mathcal{F}(M)$, the dual space is the space of densities $\mathrm{Den}(M)$. Fixing the symplectic volume $d \mu$, the $L_2$ pairing allows us to identify the dual space $\mathcal{F}^*(M)$ with  $\mathcal{F}(M)$ itself. Further, the identification $\mathfrak{z}\simeq \mathbb{R}$ implies the isomorphism $\mathfrak{z}^*\simeq \mathbb{R}$ on the level of dual spaces. Accordingly, we may consider the dual space  $ \mathcal{F}^*(M) \times \mathfrak{z}^*$ as $\mathcal{F}(M) \times \mathfrak{z}$ itself. More precisely, given $(H,c_H)$ in $ \mathcal{F}(M) \times \mathfrak{z}$, and a dual element $(f,c^*)$ in $\mathcal{F}^*(M) \times \mathfrak{z}^*$, we shall consider the pairing given by
\begin{equation}
\langle(f,c^*), (H,c_H)\rangle =c^*c_H+ \int_M fH~d \mu .
\end{equation}
Accordingly, the dual space $\mathfrak{X}_{\mathrm{ham}}^{c*}(M)$ is determined by the pairing
\begin{equation} \label{dual-calc-conf}
\begin{split}
\left\langle \Pi \otimes d \mu,X_{H}^c \right\rangle _{L_2} &= \int_{M} \langle \Pi, X_{H} - cZ
 \rangle   d \mu =\int_{M} \langle \Pi, \Omega ^\sharp(dH) - c \Omega^\sharp(\Theta)
 \rangle  d \mu
 \\&
 =\int_{M} \langle \Pi, \Omega ^\sharp(dH)\rangle d \mu -\int_{M} \langle \Pi,  c \Omega^\sharp(\Theta)
 \rangle  d \mu
 \\&
 =  -
 \int_{M} \langle \Omega ^\sharp(\Pi),  dH  
 \rangle  d \mu  -c\int_{M} \langle \Pi, \Omega ^\sharp(\Theta) \rangle d \mu
 \\&=
 -\int_{M}\iota_{\Omega ^\sharp(\Pi)} dH   d \mu  \notag +c\int_{M}  \langle \Theta, \Omega ^\sharp(\Pi) \rangle d \mu
 \\ &=
 -\int_{M}\iota_{\Omega ^\sharp(\Pi)} (dH\wedge d \mu) -\int_{M} dH\wedge \iota_{\Omega ^\sharp(\Pi)} ( d \mu) +c\int_{M} \langle \Theta, \Omega ^\sharp(\Pi) \rangle  d \mu
 \\ & =-\int_{M}dH \wedge \iota_{\Omega ^\sharp(\Pi)}
(d \mu) +c\int_{M} \langle \Theta, \Omega ^\sharp(\Pi) \rangle d \mu
\\&=
-\int_{M} d\big( H  \iota_{\Omega ^\sharp(\Pi)}d \mu\big) + \int_{M}  H ~ d \iota_{\Omega ^\sharp(\Pi)}d \mu +c\int_{M} \langle \Theta, \Omega ^\sharp(\Pi) \rangle d \mu
 \\ & =
\int_{M}H~\mathcal{L}_{\Omega ^\sharp(\Pi)} d \mu   +c\int_{M} \langle \Theta, \Omega ^\sharp(\Pi) \rangle d \mu
 \\&= \int_{M}H~ \mathrm{div} \Omega ^\sharp(\Pi)  d \mu + c\int_{M}  \langle\Theta, \Omega ^\sharp(\Pi) \rangle d \mu.
\end{split}
\end{equation} 
We thus arrive at the dual mapping of the Lie algebra homomorphism $\Phi^c$ in \eqref{Phi-c}, that is,
\begin{equation}\label{Phi-c-*}
\Phi^{c*}: \mathfrak{X}_{\rm{ham}}^{c*}(T^*M)  \longrightarrow 
\mathrm{Den}(M) \times \mathbb{R},\qquad (\Pi \otimes d \mu) \mapsto \Big( \mathrm{div} \Omega ^\sharp(\Pi)d \mu  ,\int_{M}  \langle\Theta, \Omega ^\sharp(\Pi) \rangle d \mu\Big).
\end{equation} 
In order to ensure the non-degeneracy of the pairing, therefore, we present the dual space as 
\begin{equation}
\mathfrak{X}_{\rm{ham}}^{c*}(T^*M) =\big\{\Pi \otimes d \mu \in \Lambda^1(M)\otimes \mathrm{Den}(M):  \mathrm{div} \Omega ^\sharp(\Pi) \neq 0,~ \int_{M}  \langle\Theta, \Omega ^\sharp(\Pi) \rangle d \mu  \neq 0
\big\} \cup \{0\}.
\end{equation}
This determines the following identification for the two tuple $(fd \mu,c^{*})$ with the one-form sections as
\begin{equation}\label{f-c-defn}
f(z)=\mathrm{div} \Omega ^\sharp(\Pi),\qquad c^{*}= \int_{M} \langle\Theta, \Omega ^\sharp(\Pi) \rangle d \mu. 
\end{equation}
Hence, following \eqref{coad-gen} we compute the coadjoint action of the Lie algebra $\mathfrak{X}_{\mathrm{ham}}^c(M)$ on its dual space $\mathfrak{X}_{\rm{ham}}^{c*}$ as 
\begin{equation} \label{ad*conham}
ad^*_{X_H^c}(\Pi\otimes d \mu) = \big( \mathcal{L}_{X_H^c}\Pi
+\mathrm{div}(X_H^c) \Pi \big) \otimes d \mu = \big(\mathcal{L}_{X_H^c}\Pi+nc\Pi\big) \otimes d \mu,
\end{equation}
where we have substituted the divergence of $X_H^c$ from \eqref{div-X-c}. Accordingly, in view of \eqref{LP-gen} the Lie-Poisson equation, or equivalently the coadjoint flow, is computed to be 
\begin{equation} \label{MV-conf}
\dot{\Pi}=-\mathcal{L}_{X_H^c}\Pi - cn  \Pi .
 \end{equation}
The equality \eqref{MV-conf} is the conformal kinetic equation in terms of momenta.

\textbf{Momentum-Vlasov Equation.} Let us particularly consider the case where the conformal factor is zero. In this case, one arrives at the pure Hamiltonian flow, and the Lie algebra associated with this particular case is the space $\mathfrak{X}_{\mathrm{ham}}(M)$ of Hamiltonian vector fields. Moreover, the domain of the Lie algebra homomorphism $\Phi^c$ in  \eqref{Phi-c} turns out to be the function space $\mathcal{F}(M)$, without the extension, that is
\begin{equation}\label{Phi}
 \Phi: \mathcal{F}(M)  \longrightarrow\mathfrak{X}_{\mathrm{ham}} (M) 
 ,\qquad H\mapsto X_H.
\end{equation}
The dual of this linear mapping can be computed similarly to \eqref{dual-calc-conf}. As a result, one can see that, only the first term in the dual operation $(\Phi^c)^*$ in \eqref{Phi-c-*} remains. Namely,
\begin{equation}\label{Phi-*}
\Phi^*:\mathfrak{X}_{\rm{ham}}^*(T^*M)\longrightarrow 
\mathrm{Den}(M)  ,\qquad  \Pi   \mapsto  fd \mu:=\mathrm{div} \Omega ^\sharp(\Pi)d \mu . 
\end{equation} 
This calculation allows us to define the dual space $\mathfrak{X}_{\mathrm{ham}}^*(M)$ of the space of Hamiltonian vector fields as
\begin{equation}
\mathfrak{X}_{\mathrm{ham}}^*(M): =\{\Pi
\otimes d \mu \in \Lambda ^{1}(M)\otimes \mathrm{Den}(M)):\mathrm{div}\Omega ^\sharp(\Pi)\neq 0 \} \cup {0}.
\end{equation}
Having determined the dual space properly, we can then determine the Lie-Poisson equations on $\mathfrak{X}_{\mathrm{ham}}^*(M)$. It follows from \eqref{Hamvf-Lie} that a Hamiltonian vector field is divergence-free. As such, in view of the calculation \eqref{LP}, the Lie-Poisson equation is given by
\begin{equation}\label{MV}
\dot{\Pi}= - ad^*_{X_H} \Pi = - \mathcal{L}_{X_H} \Pi,
\end{equation}
which is called the momentum-Vlasov equation in the literature \cite{EsGu12,EsSu21,Gu10}. We remark that this system is precisely equal to conformal kinetic dynamics \eqref{MV-conf} with $c$ being zero. In Appendix \ref{App-1}, we present the geometrical relationship between the momentum-Vlasov equation \eqref{MV} and the Vlasov equation which is displayed in the first line of \eqref{PV}.

 \subsection{Conformal Kinetic Dynamics in Density Formulation}

As was shown above, the dual space of the Lie algebra  $\mathcal{F}(M) \times \mathfrak{z}$ is the product $\mathrm{Den}(M) \times \mathfrak{z}$ of the space of densities with real numbers. 
Accordingly, in view of the bracket formula \eqref{alg-ext}, the coadjoint action of $\mathcal{F}(M) \times \mathfrak{z}$ on its dual $\mathrm{Den}(M) \times \mathfrak{z}$ is given by 
        \begin{equation}
        \begin{split}
        \big\langle ad^*_{(H,c_H)} (fd \mu,c^{*}),(F,c_F) \big \rangle &=  \big\langle (fd \mu,c^{*}),ad_{(H,c_H)} (F,c_F) \big \rangle 
        \\&=  \Big\langle (fd \mu, c^{*}),\big(\{H,F\}^{(S)}+c_H(Z(F)+F)-c_F(Z(H)+H),0\big)\Big \rangle
         \\&= \int_M ( f \{H,F\}^{(S)}  + f c_H(Z(F)+F)-fc_F(Z(H)+H) )d \mu.
                  \end{split}
            \end{equation}
Let us analyze the terms on the right-hand side of the last line one by one. The first term reads 
 \begin{equation}
\int_M  f \{H,F\}^{(S)} d \mu  = \int_M  \{f,H\}^{(S)}F d \mu,      
       \end{equation}       
while the third term can be written as a multiple of $c_F$. The second term, on the other hand, may be examined through
           \begin{equation}\label{egege}
        \begin{split} 
      \int_M  fc_H(Z(F)+F) d \mu &= \int_M   fc_H Z(F) d \mu+\int_M   fFc_H d \mu \\&= 
      \int_M   fc_H (\iota_ZdF) d \mu +\int_M   fFc_H  d \mu 
       \\&=  \int_M  fc_H \iota_Z ( dF\wedge d \mu) +
        \int_M   fc_H dF\wedge\iota_Z d \mu
        +\int_M   fFc_H  d \mu 
            \\&=
       \int_M   fc_H dF\wedge\iota_Z d \mu +\int_M   fFc_H d \mu \\&=       
      - \int_M   Fc_H df \wedge\iota_Z d \mu - 
       \int_M   fc_H F d\iota_Z d \mu
      + \int_M   fF c_H  d \mu \\&=       
      - \int_M   F c_H (\iota_Zdf) d \mu - 
       \int_M   fc_H F \mathrm{div}(Z) d \mu
      +\int_M   fFc_H  d \mu \\&
      =-\int_M \big(c_H Z(f) + \mathrm{div}(Z)fc_H-fc_H\big)Fd \mu.
                       \end{split}
            \end{equation}   
            Notice that, in the third line, we have used the distribution of the interior product to the wedge product of differential forms:
            \begin{equation}
                (\iota_ZdF) d \mu = \iota_Z ( dF\wedge d \mu) + dF\wedge\iota_Z d \mu . 
            \end{equation}
            See, $dF\wedge d \mu$ is identically zero since its degree is more than the dimension of the manifold. This observation reads the fourth line in \eqref{egege}. 
Now, in case $M$ is $2n$ dimensional, we recall from \eqref{divergence-Z} that the divergence of the Liouville vector field $Z$ is $\mathrm{div}(Z)=-n$. Then, 
      \begin{equation} \label{ad*-f-f}
                    ad^*_{(H,c_H)} (f,c^{*})=\Big(\{f,H\}^{(S)}-c_H Z(f) + c_H (n+1) f,-\int_M  f (Z(H)+H)d \mu \Big).
                  \end{equation} 
                  We refer to Appendix \ref{link-link} for the pure geometric link between the conformal kinetic equations \eqref{MV-conf} in momentum formulation and the conformal kinetic equation \eqref{conf-Vlasov-dens} in terms of the density. 
                  
                       The calculation \eqref{ad*-f-f} gives us the Lie-Poisson dynamics generated by a conformal vector field $X_H^c$ as a coupled PDE system 
                            \begin{equation}\label{conf-Vlasov-dens}
                                    \begin{split}
                            \frac{\partial f}{\partial t}&= \{H,f\}^{(S)}+c_H Z(f) - c_H (n+1) f,
                            \\ 
                            \frac{\partial c^{*}}{\partial t}&=\int_M  f (Z(H)+H)d \mu,
                             \end{split}
                       \end{equation}
which we call the conformal kinetic equations. 

Consider the Darboux coordinates $(q^i,p_i)$ on $M=T^*Q$, and let the Hamiltonian function be the total energy $H=p^2/2m+e\phi(q)$ of a single particle. Then the conformal kinetic equations in density formulation \eqref{conf-Vlasov-dens} take the particular form
     \begin{equation}\label{conformal-Vlasov-equation}
                                    \begin{split}
                         \frac{\partial f}{\partial t}+\frac{1}{m}\delta^{ij} p_i \frac{\partial f}{\partial q^j} -e \frac{\partial \phi}{\partial q^i}   \frac{\partial f}{\partial p_i}+ c_H(n+1)f-c_Hp_i \frac{\partial f}{\partial p_i}&=0,
                            \\ 
                            \frac{\partial c^{*}}{\partial t} -\int_M  f (
                            H- p_i \frac{\partial H}{\partial p_i})d \mu &=0.
                             \end{split} 
\end{equation} 
If the conformal factor is trivial, then the equation in \eqref{conf-Vlasov-dens} becomes  
\begin{equation}\label{Vlasov-eq}
\frac{\partial f}{\partial t}= \{H,f\}^{(S)},
\end{equation}
and, in local coordinates, we are left with the first equation in \eqref{conformal-Vlasov-equation} which turns out to be the Vlasov equation
\begin{equation}\label{Vlasov-class}
\frac{\partial f}{\partial t}+\frac{1}{m}\delta^{ij} p_i \frac{\partial f}{\partial q^j} -e \frac{\partial \phi}{\partial q^i}   \frac{\partial f}{\partial p_i}=0.
\end{equation}

\textbf{An Algebraic Route to Conformal Kinetic Equations.} Let us now note that the algebraic structure of $\mathcal{F}(M)\times\mathfrak{z}$ formulated in \eqref{alg-ext} fits the abstract formalism presented in Appendix \ref{AnAlgebraExtension}, more precisely the bracket \eqref{Alg}. Indeed, given the left action 
\begin{equation}
     \triangleright : \mathfrak{z} \times 
 \mathcal{F}(M)\longrightarrow  \mathcal{F}(M),\qquad (c,H)\mapsto c(Z(H)+H)
    \end{equation} 
of $\mathfrak{z}$ on $\mathcal{F}(M)$, and the adjoint action  
\begin{equation}
    ad_HF=\{H,F\}^{(S)}
\end{equation}
of $\mathcal{F}(M)$ on itself, the bracket \eqref{alg-ext} may be written as  
\begin{equation}
    [(H,c_H),(F,c_F)]=\big(ad_HF  +c_H\triangleright  F - c_F\triangleright H,0\big).
\end{equation}
A direct computation, then, yields the coadjoint action as
\begin{equation}\label{co-adjoint}
   ad^*_{(H,c_H)}(f,c^*)= (ad^*_Hf+f\overset{\ast}{\triangleleft}c_H, -\mathfrak{b}^*_Hf),
\end{equation}
where  
\begin{equation}
ad^*_Hf= \{f,H\}^{(S)}, \qquad f\overset{\ast}{\triangleleft}c_H=(n+1)c_Hf -c_HZ(f), \qquad \mathfrak{b}^*_Hf=\int_{M}f(Z(H)+H)d \mu.
\end{equation}
Let us note also that the coadjoint action \eqref{co-adjoint} fits exactly the one in \eqref{ad*-f-f}.

   \subsection{A Geometric Pathway to Kinetic Conformal Dynamics}\label{Sec-Geo}

In this section, we provide a geometric pathway from particle motion to irreversible kinetic dynamics. For reversible motion, this geometry has already been given in \cite{EsGrMiGu19,EsGu11,EsGu12}. Let $M$ be an $m-$dimensional manifold equipped with the local coordinates $(x^a)$, and let $\phi_t:M\rightarrow M$ denotes the flow of a vector field $X=X^a\p /\p x^a$ on $M$. 

\textbf{Complete Cotangent Lift.} Let, on the other hand, $T^*M$ be the cotangent bundle  equipped with the Darboux' coordinates $(x^a,y_a)$. The complete cotangent lift of a flow $\varphi _{t}$ on $M$ is then given by a one-parameter group of diffeomorphisms $\widehat{\varphi}_{t}$ on $T^{\ast }%
M$ satisfying%
\begin{equation}
\pi _{M}\circ \widehat{\varphi}_{t}=\varphi _{t}\circ \pi _{%
M},  \label{cotanlift}
\end{equation}%
where $\pi _{M}$ is the natural projection defined on $T^{\ast }M
$ to $M$. The vector field $%
 \widehat{X}$ on $T^*M$, which has the flow $\widehat{\varphi}_{t}$, is called the 
complete cotangent lift of $X$. We do note that, see \cite{EsGrMiGu19,EsGu11,EsGu12,yano67}, 
\begin{equation}
T\pi _{M}\circ  \widehat{X}=X\circ \pi _{M},
\end{equation}
where, in coordinates, we have
\begin{equation}\label{eq.CCL}
\widehat{X} = X^a\frac{\partial}{\partial x^a} - \frac{\partial X^b}{\partial x^a}  y_b \frac{\partial}{\partial y_a}.
\end{equation}
Let us note also that the Jacobi-Lie bracket of complete cotangent lifts is a complete cotangent lift, \cite{Marsden-67,yano67}. More precisely we have that the mapping 
\begin{equation} 
\widehat{} ~:\mathfrak{X}(M) \longrightarrow \mathfrak{X}%
\left( T^{\ast }M\right), \qquad X\mapsto \widehat{X}  \label{isos}
\end{equation}
is a one-to-one Lie algebra homomorphism. 

\textbf{Divergence Lift.} Let $M$ be a (volume) manifold, and let $Z:=\Omega^\sharp(\Theta_M)$ be the Liouville vector field on the cotangent bundle $T^*M$, where $\Theta_M$ is the canonical one-form on $T^*M$. Moreover, let also $\pi^*_M\mathcal{F}(M)$ be the pullback of the space of functions $\mathcal{F}(M)$, to the level of cotangent bundle, using the cotangent bundle projection $\pi_M: T^*M\mapsto M$. Now, since the canonical Poisson bracket on the cotangent bundle $T^*M$ vanishes on $\pi^*_M\mathcal{F}(M)$, it turns out to be a Lie subalgebra of the space $\mathcal{F}(T^*M)$ of smooth functions on $T^*M$. We then define the space 
\begin{equation}
\mathfrak{w}=\{FZ\in \mathfrak{X}(T^*M): F\in\pi^*_M\mathcal{F}(M) \},
\end{equation}
which evidently is a Lie subalgebra of $\mathfrak{X}(T^*M)$, as the Jacobi-Lie bracket of such vector fields vanishes. Let us remark that in Darboux' coordinates $(x^a,y_a)$, the Liouville vector field $Z$ and an arbitrary element $FZ$ take the form of
\begin{equation}
Z= - y_a\frac{\partial }{\partial y_a},\qquad FZ=- F(x^a)y_a\frac{\partial }{\partial y_a}.
\end{equation}

Next, let $X$ be a vector field on $M$ with possibly a non-zero divergence. Since $\mathrm{div}(X)$ is a function on $M$, we can lift it to the cotangent bundle by means of the Liouville vector field $Z$ as
\begin{equation}\label{div-lift}
\mathcal{D}:\mathfrak{X}(M)\longrightarrow   \mathfrak{X}(T^*M),\qquad X\mapsto \mathrm{div}(X)Z.
\end{equation}
A direct computation reads that for arbitrary vector fields $X$ and $Y$, we have that 
\begin{equation}
[\mathcal{D}(X),\mathcal{D}(Y)]= 0 .
\end{equation}
This means that the image space of $\mathcal{D}$ operator reads a trivial Lie subalgebra of $\mathfrak{X}(T^*M)$. 
Recalling the identity 
\begin{equation}
\mathrm{div}[X,Y]=\mathrm{div}(X)(Y)-\mathrm{div}(Y)(X),
\end{equation}
we see that the mapping $\mathcal{D}$ turns out to be a Lie algebra homomorphism only if we restrict the domain to the space $\mathfrak{X}_c(M)$ of vector fields with constant divergences. The kernel of $\mathcal{D}$, on the other hand, happens to be the space of divergence-free vector fields. 

Now we collect the complete cotangent lift  \eqref{isos} and the divergence lift \eqref{div-lift} to define a mapping from the space $\mathfrak{X}_c(M)$ of vector fields with constant divergence into the space $\mathfrak{X}(T^*M)$ of  vector fields as
\begin{equation}\label{kappa}
\kappa:\mathfrak{X}_c(M)\longrightarrow   \mathfrak{X}(T^*M),\qquad X\mapsto \widehat{X}+\mathrm{div}(X)Z.
\end{equation}
A direct calculation proves that the Jacobi-Lie bracket of a complete cotangent lifts $\widehat{X}$ and a divergence lift $\mathcal{D}(X)$ is trivial. As such, the mapping $\kappa$ is a Lie algebra homomorphism. Moreover, in Darboux coordinates, the image of a vector field under $\kappa$ is computed to be
\begin{equation}\label{eq.Xi}
\kappa(X)= X^a\frac{\partial}{\partial x^a} - \Big(\frac{\p X^b} {\p x^b} y_a + \frac{\partial X^b}{\partial x^a}  y_b \Big)\frac{\partial}{\partial y_a}.
\end{equation} 
If the vector field $X$ is divergence-free then we are left only with the complete cotangent vector field. 

\textbf{Holonomic Part.}
Once again $\pi_M: T^*M\mapsto M$ being the cotangent bundle, let $J^{1}T^*M$ be the first jet bundle (which happens to be a $2n+n^2$ dimensional manifold) with the induced coordinates 
\begin{equation}
(x,y,y_{x})=(x^a,y_a,\p y_b / \p x^a).
\end{equation}
Let $X$ be a vector field on $M$, and $\sigma$ a one-form. The Lie derivative (directional derivative) of a smooth function $F$, defined on the total space $T^*M$, with respect to the vector field $X$ can be computed by means of $\sigma$ as $\mathcal{L}_X(F\circ \sigma)$. Accordingly, the definition of the holonomic lift $X^{hol}$ of the vector field $X$ may be given by the identity 
\begin{equation} \label{hol-lift}
X^{hol}(F)\circ \sigma:=\mathcal{L}_X(F\circ \sigma),
\end{equation}
see for instance \cite{olver86,Saunders-book}. In local coordinates, the holonomic lift of $U=U^{a}{\partial }/{\partial x^{a}}$ is computed to be
\begin{equation}
U^{hol}=U^{a}\frac{\partial }{\partial x^{a}}+U^{a}\frac{\p y_b }{ \p x^a}\frac{\partial 
}{\partial y_b}.
\end{equation}
We note that $U^{hol}$ is not a classical vector field, since its coefficients depend on the first-order jet bundle. Such kinds of sections are called the generalized vector fields, \cite{Kosmann-80}. 

For a projectable vector field $Y$ on the cotangent bundle $T^*M$, the holonomic part $HY$ is defined to be the holonomic lift of its projection, that is, for a vector field $Y=Y^{a}(x)\partial /\partial x^{a}+Y_b(x,y)\partial /\partial
y_b$, 
\begin{equation} \label{YtoHY}
HY:=(T\pi\circ Y)^{hol}=Y^{a}\frac{\partial }{\partial x^{a}}+Y^{a}\frac{\p y_b }{ \p x^a}\frac{\partial 
}{\partial y_b}.
\end{equation}

A generalized vector field on $T^*M$, then, is of the form 
\begin{equation}
\chi =\chi  ^{a}\left( x\right)   \frac{\partial }{\partial x^{a}}%
 +\chi_b\left( x,y,y_{x}\right)  \frac{%
\partial }{\partial y_b}.
\label{genvec}
\end{equation}%
As such, the first order prolongation $pr^{1}\chi $ of $\chi$ may is given by 
\begin{equation}
pr^{1}\chi =\chi +\Delta _{ba}\frac{\partial }{\partial (\p y_b / \p x^a)},\qquad \Delta _{ba}=D_{x^{a}}\big( \chi_e-\chi^b (\p y_e / \p x^b)\big)
+(\p^2 y_e / \p x^a \p x^b)\chi^{b},
\end{equation}%
where $D_{x^{a}}$ is the total derivative operator with respect to $x^{a}$,
and $\p^2 y_e / \p x^a \p x^b$ is an element of the second order jet bundle. Furthermore, the Lie bracket of two first-order generalized vector fields $\chi $ and $\psi$ is
the unique first-order generalized vector field is given by
\begin{equation}
\left[ \chi  ,\psi \right] _{pro}=\left( pr^{1}\chi  \left( \psi ^{a}\right)
-pr^{1}\psi\left( \chi ^{a}\right) \right) \frac{\partial }{\partial x^{a}}%
+\left( pr^{1}\chi  \left( \psi_a\right) -pr^{1}\psi \left( \chi 
_a\right) \right) \frac{\partial }{\partial y_a}.
\label{Liepro}
\end{equation}%
Accordingly, the holonomic part operation $H: Y\rightarrow HY$ defined in \eqref{YtoHY} is a Lie algebra homomorphism from the space of projectable vector fields into the space of generalized vector fields of order one, equipped with the bracket \eqref{Liepro}. 

Finally, the holonomic part of the image space of the mapping $\kappa$ given by \eqref{eq.Xi} is computed to be
\begin{equation}\label{HXi}
H\kappa(X)= X^a\frac{\partial}{\partial x^a} +X^{a}\frac{\p y_b }{ \p x^a}\frac{\partial 
}{\partial y_b}.
\end{equation} 
We remark that the term $\mathrm{div}(X)Z$ does not contribute to the holonomic part since it is a pure vertical vector and that the composition mapping $X\mapsto H\kappa(X)$ is a Lie algebra homomorphism since both \eqref{kappa} and \eqref{YtoHY} are Lie algebra homomorphisms. 

\bigskip 

\textbf{Vertical Representative.} The holonomic lift operation $U^{hol}$ copies the dynamics on the base manifold to the cotangent bundle in terms of the action of the vector field $U$ on the fiber coordinates. Hence, the vertical motion (that is, the dynamics governing the sections) is obtained by subtracting the holonomic part $HY$ from a projectable vector field $Y$ on $T^*M$, \cite{olver86}. In short, we shall call 
\begin{equation}
VY=Y-HY=\left( Y^{\alpha
}-Y^{a}u_{a}^{\lambda }\right) \frac{\partial }{\partial u^{\lambda }}
\end{equation}
the vertical representative of $Y$. We note at once that $VY$ lies in the kernel of $T\pi_M$. 

The vertical representative of $\kappa(X)$ defined in \eqref{eq.Xi} is computed to be 
\begin{equation}\label{VXi}
 V\kappa(X)=-\Big(\frac{\p X^b} {\p x^b} y_a + \frac{\partial X^b}{\partial x^a} y_b + X^{b}\frac{\p y_a }{ \p x^b}  \Big)\frac{\partial}{\partial y_a}.
 \end{equation}
Given the one-form $\Pi=y_adx^a$, a quick computation yields
 \begin{equation}
\dot{\Pi}=-\mathcal{L}_X(\Pi)-\mathrm{div}(X)\Pi,
  \end{equation}
  that is, the local formulation of the vertical representative $V\kappa(X)$ is exactly the Lie-Poisson dynamics \eqref{LP-gen}. Furthermore, the vertical representative operation $\kappa(X)\mapsto  V\kappa(X)$ is a Lie algebra homomorphism, endowing the image space with the prolonged bracket \eqref{Liepro}.

To sum up, we have the following Lie algebra homomorphisms  for the particle motion to the motion of the continuum: 
\begin{equation} 
\xymatrix{
{\begin{array}{c} \text{Particle}  \\ \text{motion}, ~ X \end{array}}\ar[rrr]^{\kappa~in~\eqref{kappa}}
&&&{\begin{array}{c} \text{Lifted} \\ \text{motion}, ~ \kappa(X) \end{array}}\ar[rrr]^{V~in~\eqref{VXi}}
&&&
{\begin{array}{c} \text{Kinetic motion} \\ \text{of fibers}, V\kappa(X). \end{array}}
}
  \end{equation} 
We do note that this geometric path provides the geometrization of the conformal kinetic dynamics in \eqref{MV-conf}. For the autonomous cases, one reduces the mapping $\kappa$ to the cotangent bundle and arrives at the momentum Vlasov dynamics in \eqref{MV}.

\section{Contact Kinetic Theory}\label{Sec:kin-con}


Hamiltonian kinetic theory can be geometrically constructed through the following steps. First, the Lie group of transformations is determined. Associated with that Lie group, the Lie algebra and its dual are obtained. On the Lie algebra dual, there exists the Lie-Poisson bracket. Finally, once energy is specified, the Lie-Poisson bracket and energy yield the evolution equation for the distribution function. In this section, we follow these steps using contact geometry (that is contact diffeomorphism group) to derive the kinetic motion of a collection of particles, each obeying contact Hamiltonian motion individually.

\subsection{Contact Manifolds}

Let $\bar{M}$ be an odd, say $(2n + 1)$, dimensional manifold. A contact structure on $\bar{M}$ is a maximally non-integrable smooth distribution of codimension one, and it is locally given by the kernel of a one-form $\eta$ such that 
\begin{equation}
    d\eta^n
\wedge \eta \neq 0.
\end{equation}
Such a one-form $\eta$ is called a (local) contact form \cite{arnold1989mathematical,LiMa87}. In the present manuscript, we shall consider the existence of a global contact one-form. A contact one-form for a given contact structure is not unique. Indeed, if $\eta$ is a contact one-form for a fixed contact structure, then $\lambda \eta$ also defines the same contact structure for any non-zero real-valued function $\lambda$ defined on $\bar{M}$. In short, we call a  $(2n+1)-$dimensional manifold  $\bar{M}$ as a contact manifold if it is equipped with a contact one-form $\eta$ satisfying $d\eta^n
\wedge \eta \neq 0$, and we shall denote a contact manifold by $(\bar{M}, \eta)$. 

Given a contact one-form $\eta$, the vector field $\mathcal{R}$ satisfying 
\begin{equation}
\iota_{\mathcal{R}}\eta =1,\qquad \iota_{\mathcal{R}}d\eta =0
\end{equation}
is unique, and it is called the Reeb vector field. There is, on the other hand, a musical isomorphism $\flat$ from the space of sections of the tangent bundle $T\bar{M}$ to the sections of the cotangent bundle $T^*\bar{M}$ defined by 
\begin{equation}\label{flat-map}
\flat:\mathfrak{X}(\bar{M})\longrightarrow \Lambda^1(\bar{M}),\qquad Y\mapsto \iota_Yd\eta+\eta(Y)\eta.
\end{equation} 
It is worth noting that the image of the Reeb vector field $\mathcal{R}$ under the musical mapping is the contact one-form $\eta$. We shall denote the inverse of \eqref{flat-map} by $\sharp$. Referring to this, we define a bivector field $\Lambda$ on  $M$ as
\begin{equation}\label{Lambda}
\Lambda(\alpha,\beta)=d\eta(\sharp\alpha, \sharp \beta). 
\end{equation}
Referring to the bivector field $\Lambda$ we introduce the following musical mapping 
\begin{equation}\label{Sharp-Delta}
\sharp_\Lambda: \Lambda^1(\bar{M}) \longrightarrow \mathfrak{X}(\bar{M}), \qquad  \alpha\mapsto \Lambda(\bullet,\alpha)= \sharp \alpha - \alpha(\mathcal{R})  \mathcal{R}. 
\end{equation}
The kernel of the mapping $\sharp_\Lambda$ is spanned by the contact one-form $\eta$  so it fails to be an isomorphism. A contact manifold $(\bar{M},\eta)$ admits a Jacobi manifold structure \cite{Lichnerowicz-Jacobi,Marle-Jacobi}, see \cite{Bruce17} for a more recent exposition. This realization permits us to define a Jacobi (contact) bracket 
\begin{equation}\label{cont-bracket-L} 
\{\bar{F},\bar{H}\}^{(C)} =\Lambda(d\bar{F},d\bar{H}) +\bar{F}\mathcal{R}(\bar{H}) - \bar{H}\mathcal{R}(\bar{F}).
\end{equation}
See that the bracket satisfies the Jacobi identity but the Leibniz's identity is violated due to the last two terms (the Reeb terms) on the right-hand side.

\textbf{Contact Hamiltonian Motion.} 
For a Hamiltonian function $\bar{H}$ on a contact manifold $(\bar{M},\eta)$, there is a corresponding contact Hamiltonian vector field $\xi_{\bar{H}}$ given by
\begin{equation}
\iota_{\xi_{\bar{H}}}\eta =-\bar{H},\qquad \iota_{\xi_{\bar{H}}}d\eta =d\bar{H}-\mathcal{R}(\bar{H}) \eta,   \label{contact}
\end{equation}
where $\mathcal{R}$ is the Reeb vector field, and $\bar{H}$ is called the contact Hamiltonian function \cite{bravetti2017contact,de2019contact,esen2021contact}. We also have
\begin{equation}\label{3-def}
\flat(\xi_{\bar{H}})=d\bar{H}-(\mathcal{R}(\bar{H})+\bar{H})\eta.
\end{equation}%

\textbf{Dissipation.}
Let us denote a contact Hamiltonian system as a three-tuple $(\bar{M},\eta,\bar{H})$, where $(\bar{M},\eta)$ is a contact manifold and $\bar{H}$ is a smooth real function on $\bar{M}$. A direct computation determines a conformal factor for a given contact vector field via
\begin{equation}\label{L-X-eta}
\mathcal{L}_{\xi_{\bar{H}}}\eta =
d\iota_{\xi_{\bar{H}}}\eta+\iota_{\xi_{\bar{H}}}d\eta= -\mathcal{R}(\bar{H})\eta.
\end{equation}
According to \eqref{L-X-eta}, the flow of a contact Hamiltonian system preserves the contact structure, but it preserves neither the contact one-form nor the Hamiltonian function. Instead, we obtain
\begin{equation}
{\mathcal{L}}_{\xi_{\bar{H}}} \, \bar{H} = - \mathcal{R}(\bar{H}) \bar{H}.
\end{equation}
Being a non-vanishing top-form we may consider $  d\eta^n
\wedge \eta$ as a volume form on $\bar{M}$. 
The Hamiltonian motion does not preserve the volume form since
\begin{equation}
{\mathcal{L}}_{\xi_{\bar{H}}}  \, (d\eta^n
\wedge \eta) = - (n+1)  \mathcal{R}(\bar{H}) d\eta^n
\wedge \eta.
\end{equation}
Assuming the dimension of $\bar{M}$ to be $2n+1$, we compute the divergence of a contact vector field as
\begin{equation} \label{div-X-H}
\mathrm{div}(\xi_{\bar{H}})= -  (n+1)  \mathcal{R}(\bar{H}).
\end{equation}
However, it is immediate to see that, for a nowhere vanishing Hamiltonian function $\bar{H}$, the quantity $ {\bar{H}}^{-(n+1)}   (d\eta)^n \wedge\eta$ 
is preserved along the motion (see \cite{BrLeMaPa20}).

A direct computation proves that the contact bracket and the contact vector field are related as
\begin{equation}
\begin{split}
\{\bar{F},\bar{H}\}^{(C)} &=- \iota_{[\xi_{\bar{H}},\xi_{\bar{F}}]}\eta =-\mathcal{L}_{\xi_{\bar{H}}}\iota_{\xi_{\bar{F}}} \eta+ 
\iota_{\xi_{\bar{F}}}\mathcal{L}_{\xi_{\bar{H}}} \eta
\\&=-\mathcal{L}_{\xi_{\bar{H}}}(-\bar{F}) + \iota_{\xi_{\bar{F}}}(-\mathcal{R}(\bar{H})\eta)
\\&= \xi_{\bar{H}}(\bar{F})+\bar{F}\mathcal{R}(\bar{H}).
\end{split} 
\end{equation}
This observation is important. We remark that the flow generated by the contact vector field and the flow generated by the contact bracket are not the same.

 \textbf{Darboux Coordinates and Contactization of a Symplectic Manifold.} We start with an exact symplectic manifold $(M,\Omega=-d\Theta)$ and consider  the principal line bundle
\begin{equation}
(\bar{M},\eta )\overset{pr}{\longrightarrow }%
(M,\Omega=-d\Theta). 
\end{equation}%
For a local coordinate system $z$ on the circle (which we shall consider being $\mathbb{R}$), and the Darboux coordinates $(q^i,p_i)$ on the symplectic manifold, the contact manifold admits the Darboux coordinates $(q^i,p_i,z)$ on $\bar{M}$. In this realization, the contact one-form and the associated Reeb field are 
\begin{equation}\label{eta-Q}
    \eta=dz-\bar{\Theta}=dz-p_idq^i,\qquad \mathcal{R}=\frac{\partial}{\partial z},
\end{equation}
where $\bar{\Theta}$ is the pullback of the potential one-form on $M$. This suggests the coordinates $(q^i,p_i,z)$ on the contact manifold $\bar{M}$. In this case, the volume form $\overline{d \mu}$ on the contact manifold is computed to be 
\begin{equation}\label{contact-volume}
\overline{d \mu}=dz\wedge d \mu
\end{equation}
where $d \mu$ is the symplectic volume in \eqref{symp-volume}. 
The bivector $\Lambda$ in \eqref{Lambda} is computed to be 
\begin{equation}\label{AP-Bra}
\Lambda= \frac{\partial  }{\partial q^{i}}\wedge \frac{\partial  }{\partial p_{i}} + p_i \frac{\partial  }{\partial z} \wedge \frac{\partial  }{\partial p_i}. 
\end{equation} 
 In terms of the Darboux coordinates, we compute the musical mapping $\sharp_\Lambda$
  in \eqref{Sharp-Delta}   as
\begin{equation}\label{zap}
\sharp_\Lambda:\alpha_i dq^i + \alpha^i dp^i + 
u dz\mapsto 
\alpha^i \frac{\partial}{\partial q^i}-(\alpha_i + p_i u)\frac{\partial}{\partial p_i}  + \alpha^ip_i  \frac{\partial}{\partial z}. 
\end{equation}
Then, in this local picture, the contact bracket \eqref{cont-bracket-L} is 
\begin{equation}\label{Lag-Bra}
\{\bar{F},\bar{H}\}^{(C)} = \frac{\partial \bar{F}}{\partial q^{i}}\frac{\partial \bar{H}}{\partial p_{i}} -
\frac{\partial \bar{F}}{\partial p_{i}}\frac{\partial \bar{H}}{\partial q^{i}} + \left(\bar{F}  - p_{i}\frac{\partial \bar{F}}{\partial p_{i}} \right)\frac{\partial \bar{H}}{\partial z} -
\left(\bar{H}  - p_{i}\frac{\partial \bar{H}}{\partial p_{i}} \right)\frac{\partial \bar{F}}{\partial z}.
\end{equation}

A symplectic manifold 
$M$ locally looks like a cotangent manifold. Accordingly, without loss of generalization, one may substitute the symplectic manifold $M$ with the cotangent bundle $T^*Q$. In this case, the contact manifold $\bar{M}$ locally turns out to be the extended cotangent bundle $T^*Q\times \mathbb{R}$.

\subsection{Dynamics on Contact Manifolds}

This subsection introduces two different dynamical vector fields that can be determined on a contact manifold $(\bar{M},\eta)$. To have these realizations, for a given Hamiltonian function $\bar{H}$, we first recall the 
contact Hamiltonian vector field definition in \eqref{3-def} and write it as
\begin{equation}\label{defn-3}
\xi_{\bar{H}}=\sharp(d\bar{H})-\mathcal{R}(\bar{H})\mathcal{R} - \bar{H}\mathcal{R}. 
\end{equation}
As we depict in the sequel, the space of such vector fields determines a Lie algebra as a manifestation of the Jacobi manifold structure of  the contact manifold. By using only the first and third terms on the right-hand side we define a strict contact Hamiltonian vector field as
\begin{equation}
Y_{\bar{H}}=\sharp(d\bar{H})- \bar{H}\mathcal{R}.
\end{equation}
Let us now depict all the algebraic properties of these dynamics in detail.

\textbf{Contact Diffeomorphisms and Contact Hamiltonian Vector Fields.}
For a contact manifold $(\bar{M},\eta)$, a contact diffeomorphism (contactomorphism) is the one that preserves the contact structure. We  denote the
group of contact diffeomorphisms by \cite{banyaga97} 
\begin{equation}\label{Diff-con}
{\rm Diff}_{\mathrm{con}} ( \bar{M} ) =\left\{ \varphi \in {\rm Diff} ( \bar{M}  ) :\varphi ^{\ast }\eta =\gamma  \eta, \quad \gamma  \in \mathcal{F}%
( \bar{M} ) \right\} .
\end{equation}
Here, ${\rm Diff} ( \bar{M} )$ is standing for the group of all diffeomorphism on $\bar{M} $. Notice that the existence of $\gamma $ in the definition manifests the conformal definition of the contact structure. 
A vector field on the contact
manifold $( \bar{M},\eta) $ is a contact vector field (called also infinitesimal conformal contact diffeomorphism) if
it generates a one-parameter group of contact diffeomorphisms. Accordingly, the space of contact vector fields is given by
\begin{equation}\label{algcon}
\mathfrak{X}_{\mathrm{con}} ( \bar{M}  ) =\left\{ X\in \mathfrak{X} ( 
\bar{M} ) :\mathcal{L}_{X}\eta  =-\lambda \eta ,\quad \lambda
\in \mathcal{F} ( \bar{M}  ) \right\} .  
\end{equation}
Sometimes a contact vector field is denoted by a two-tuple $(X,\lambda)$ to exhibit the conformal factor $\lambda$. 
A direct observation reads from \eqref{L-X-eta}  that a contact Hamiltonian vector field $\xi_{\bar{H}}$ is a contact vector field with conformal parameter $\lambda=\mathcal{R}( \bar{H})$ so it belongs to $\mathfrak{X}_{\mathrm{con-ham}} ( \bar{M}  )$, for more details, see \cite{Br17,bravetti2017contact,de2021hamilton,de2019contact}. In this work, our interest is the space of contact Hamiltonian vector fields
\begin{equation}\label{X-con}
\mathfrak{X}_{\mathrm{con-ham}} ( \bar{M} )=\big\{\xi_{\bar{H}}\in \mathfrak{X}(\bar{M} ): \iota_{\xi_{\bar{H}}}\eta =-\bar{H},~ \iota_{\xi_{\bar{H}}}d\eta =d\bar{H}-\mathcal{R}(\bar{H}) \eta\big\}.
\end{equation}
This space is a Lie subalgebra of space of all vector fields as a manifestation of the identity  
\begin{equation} 
\left[\xi_{\bar{F}},\xi_{\bar{H}}\right]=-\xi_{ \{ \bar{F},\bar{H} \}^{(C)}}.  
\end{equation}
So, one may establish the following
isomorphism from the space of real smooth functions on $\bar{M}$ to the space of contact Hamiltonian vector fields
\begin{equation} \label{iso}
\Psi: ( \mathcal{F}( \bar{M}
) ,\left\{ \bullet,\bullet \right\} ^{(C)})\longrightarrow 
\left( \mathfrak{X}_{\mathrm{con-ham}} ( \bar{M} ) ,-\left[\bullet ,\bullet
\right] \right) , \qquad \bar{H} \mapsto \xi_{\bar{H}}.
\end{equation}
Referring to the Darboux's coordinates $(q^{i},p_{i},z)$, for a function $\bar{H}=\bar{H}(q^i,p_i,z)$, the contact Hamiltonian vector field determined in \eqref{contact} becomes
\begin{equation}\label{con-dyn}
\xi_{\bar{H}}=\frac{\partial \bar{H}}{\partial p_{i}}\frac{\partial}{\partial q^{i}}  - \left(\frac{\partial \bar{H}}{\partial q^{i}} + \frac{\partial \bar{H}}{\partial z} p_{i} \right)
\frac{\partial}{\partial p_{i}} + \left(p_{i}\frac{\partial \bar{H}}{\partial p_{i}} - \bar{H}\right)\frac{\partial}{\partial z}.
\end{equation}
Thus, we obtain the contact Hamilton's equations for $\bar{H}$ as
\begin{equation}\label{conham}
\frac{dq^{i}}{dt} = \frac{\partial \bar{H}}{\partial p_{i}}, 
\qquad
 \frac{dp_{i}}{dt}  = -\frac{\partial \bar{H}}{\partial q^{i}}- 
p_{i}\frac{\partial \bar{H}}{\partial z}, 
\qquad \frac{dz}{dt} = p_{i}\frac{\partial \bar{H}}{\partial p_{i}} - \bar{H}.
\end{equation}
In particular, the Reeb vector field becomes $\mathcal{R}=\partial/\partial z$. The divergence of a contact Hamiltonian vector field \eqref{div-X-H} is then
\begin{equation} \label{div-X-H-}
\mathrm{div}(\xi_{\bar{H}})= -  (n+1)  \mathcal{R}(\bar{H}) =  - (n+1) \frac{\partial \bar{H}}{\partial z}.
\end{equation}
Contact dynamics finds many applications in various fields of physics especially in thermodynamics see, for example, \cite{Bravettithermo,Goto15,Grmela14,Mrugala,Simoes-thermo}. 

 \textbf{Quantomorphisms and Strict Contact Hamiltonian Vector Fields.} Let us consider a contact manifold $(\bar{M},\eta)$.  By asking the conformal factor $\gamma$ to be the unity for a contact diffeomorphism in \eqref{Diff-con}, one arrives at the conservation of the contact form $\varphi^*\eta = \eta$. We call such a mapping as a strict contact diffeomorphism (or a quantomorphism). For a contact manifold $(\bar{M},\eta)$, 
we denote the group of all strict contact diffeomorphisms  as
\begin{equation}
{\rm Diff}_{\mathrm{st-con}} ( \bar{M})
 =\left\{ \varphi \in {\rm Diff} ( \bar{M}) :\varphi ^{\ast }\eta =   \eta \right\} \subset {\rm Diff}_{\mathrm{con}} ( \bar{M})
.
\end{equation}
The Lie algebra of this group consists of so-called strict contact vector fields (or, infinitesimal quantomorphisms, or infinitesimal strict contact diffeomorphism)
\begin{equation}\label{X-st-con}
\mathfrak{X}_{\mathrm{st-con}}  ( \bar{M}) =\left\{ X \in \mathfrak{X} ( \bar{M} ) :\mathcal{L}_{X}\eta =0\right\} \subset \mathfrak{X}_{\mathrm{\mathrm{con}} }( \bar{M} ). 
\end{equation}
Notice that for a given Hamiltonian function $\bar {H}$, the contact Hamiltonian vector field $\xi_H$ defined in \eqref{contact} is a strict contact vector field if and only if $d\bar{H}(\mathcal{R})=0$. 
This reads the following space of strict contact Hamiltonian vector fields
\begin{equation}\label{X-st-con-ham}
\mathfrak{X}_{\mathrm{st-con-ham}} ( \bar{M} )=\big\{Y_{\bar{H}}\in \mathfrak{X}(\bar{M} ): \iota_{Y_{\bar{H}}}\eta =-\bar{H},~ \iota_{Y_{\bar{H}}}d\eta =d\bar{H} \big\} \subset \mathfrak{X}_{\mathrm{con-ham}} ( \bar{M} ).
\end{equation}
Referring to the local realization in \eqref{eta-Q} given in terms of the Darboux coordinates $(q^i,p_i,z)$, it is possible to see that  to generate a strict contact Hamiltonian vector field, a function $\bar{H}$ must be independent of the fiber variable $z$. 
For two functions, those that are not dependent on the fiber variable $z$, the contact bracket $\{\bullet,\bullet \}^{(C)}$ in \eqref{Lag-Bra} locally turns out to be equal to the canonical Poisson bracket on the symplectic manifold $M$. Accordingly, a direct calculation reads that 
\begin{equation}
\left[ Y_{\bar{H}},Y_{\bar{F}}\right]=-Y_{ \{ \bar{H},\bar{F} \} ^{(C)}}.
\end{equation}
Note that, one has the following identities in terms of the musical mapping $\flat$ in \eqref{flat-map} and its inverse $\sharp$ as 
\begin{equation}
\flat(Y_{\bar{H}})=  d\bar{H} - \bar{H} \eta, \qquad Y_{\bar{H}}=\sharp(d \bar{H}) -\bar{H} \mathcal{R}.
\end{equation}

Referring to the Darboux's coordinates $(q^{i},p_{i},z)$, for a Hamiltonian function $\bar{H}=\bar{H}(q^i,p_i)$ independent of the fiber variable $z$, the strict contact Hamiltonian vector field is 
\begin{equation}\label{st-con-dyn}
Y_{\bar{H}}=\frac{\partial \bar{H}}{\partial p_{i}}\frac{\partial}{\partial q^{i}}  -  \frac{\partial \bar{H}}{\partial q^{i}} 
\frac{\partial}{\partial p_{i}} + \left(p_{i}\frac{\partial \bar{H}}{\partial p_{i}} - \bar{H}\right)\frac{\partial}{\partial z}.
\end{equation}
Thus, we obtain strict contact Hamilton's equations as 
\begin{equation}\label{st-conham}
\frac{dq^{i}}{dt} = \frac{\partial \bar{H}}{\partial p_{i}}, 
\qquad
 \frac{dp_{i}}{dt}  = -\frac{\partial \bar{H}}{\partial q^{i}}, 
\qquad \frac{dz}{dt} = p_{i}\frac{\partial \bar{H}}{\partial p_{i}} - \bar{H}.
\end{equation}
See that this flow is divergence-free.

\subsection{Kinetic Dynamics in Terms of Momenta}

Let $\bar{M}$ be the extended cotangent bundle with the contact one-form $\eta$. We shall now determine the kinetic motion of contact particles. To this end, we shall lift the particle motion. 

\textbf{The Dual Space of Contact Hamiltonian Vector Fields.} 
Let us now determine the dual space $\mathfrak{X}_{\mathrm{con-ham}}^* (\bar{M})$ of the space contact vector fields $\mathfrak{X}_{\mathrm{con-ham}}(\bar{M})$ given in \eqref{X-con}. We first note that   $\mathfrak{X}_{\mathrm{con-ham}}^*(\bar{M})$ is a subspace of the space $\Lambda^1(\bar{M})\otimes \mathrm{Den} (\bar{M})$ of one-form densities. To be more precise, we compute the $L_2$-pairing (simply multiply-and-integrate) of an arbitrary contact vector field $\xi_{\bar{H}}$ with a one-form density $\bar{\Pi}\otimes \overline{d \mu}$. Making use of the identities of the Cartan calculus, we obtain
\begin{equation}\label{dual-1}
\begin{split}
 \langle \bar{\Pi} \otimes \overline{d \mu},& \xi_{\bar{H}}  \rangle _{L_2} = \int \langle \bar{\Pi}, \xi_{\bar{H}} 
 \rangle   \overline{d \mu} =\int \langle \bar{\Pi}, \sharp(d\bar{H}) - (\mathcal{R}(\bar{H})+\bar{H} )\mathcal{R}
 \rangle  \overline{d \mu} 
\\
&= - \int \langle \sharp\bar{\Pi}, d\bar{H}  \rangle  \overline{d \mu}  + \int (\mathcal{R}(\bar{H})+\bar{H} ) \langle \sharp\bar{\Pi},\eta  \rangle  \overline{d \mu}  
\\
&= - \int \big(\iota_{\sharp\bar{\Pi}}  d\bar{H} \big)  \overline{d \mu}  + \int  \mathcal{R}(\bar{H})  \langle \sharp\bar{\Pi},\eta  \rangle  \overline{d \mu}  + \int   \bar{H}   \langle \sharp\bar{\Pi},\eta  \rangle  \overline{d \mu} 
 \\&=  -
 \int  d\bar{H} \iota_{\sharp\bar{\Pi}} \overline{d \mu} + 
 \int  \iota_{\mathcal{R}}d\bar{H}   \langle \sharp\bar{\Pi},\eta  \rangle  \overline{d \mu} + \int   \bar{H}   \langle \sharp\bar{\Pi},\eta  \rangle  \overline{d \mu} 
  \\&=  \int  \bar{H} d \iota_{\sharp\bar{\Pi}} \overline{d \mu}
  + \int \langle \sharp\bar{\Pi},\eta  \rangle d\bar{H} \wedge  \iota_{\mathcal{R}} \overline{d \mu} + \int   \bar{H}   \langle \sharp\bar{\Pi},\eta  \rangle  \overline{d \mu} 
   \\&= \int  \bar{H} d \iota_{\sharp\bar{\Pi}} \overline{d \mu}
  - \int \bar{H} d\langle \sharp\bar{\Pi},\eta  \rangle \wedge  \iota_{\mathcal{R}} \overline{d \mu} 
  -\int \bar{H}  \langle \sharp\bar{\Pi},\eta  \rangle    d\iota_{\mathcal{R}} \overline{d \mu}
  + \int   \bar{H}   \langle \sharp\bar{\Pi},\eta  \rangle  \overline{d \mu}   
  \\ &= 
  \int  \bar{H} \Big( \mathrm{div}(\sharp\bar{\Pi}) \overline{d \mu}
  -   \iota_{\mathcal{R}}d\langle \sharp\bar{\Pi},\eta  \rangle  - \langle \sharp\bar{\Pi},\eta  \rangle    \mathrm{div}(\mathcal{R})    + \langle \sharp\bar{\Pi},\eta  \rangle \Big)  \overline{d \mu}    
 \\ &= 
  \int  \bar{H} \Big( \mathrm{div}(\sharp\bar{\Pi}) 
  -   \mathcal{L}_\mathcal{R} \langle \sharp\bar{\Pi},\eta  \rangle       + \langle \sharp\bar{\Pi},\eta  \rangle \Big)  \overline{d \mu},
\end{split}
\end{equation}
where $\mathrm{div}$ stands for the divergence with respect to the contact volume $\overline{d \mu}$ in \eqref{contact-volume}. Accordingly, once the volume form is fixed, the non-degeneracy of the pairing motivates us to define the dual space as
\begin{equation} 
\mathfrak{X}_{\mathrm{con-ham}}^* (\bar{M}) = \big \{\bar{\Pi}  \in 
\Lambda^1 (\bar{M}) : \mathrm{div}(\sharp\bar{\Pi}) 
  -   \mathcal{L}_\mathcal{R} \langle \sharp\bar{\Pi},\eta  \rangle     
  + \langle \sharp\bar{\Pi},\eta  \rangle\neq 0 \big
\}  \cup \{0\}.
\end{equation}
We next recall the Lie algebra isomorphism $\bar{H}\mapsto \xi_{\bar{H}}$ of \eqref{iso}. Identifying the dual $\mathcal{F}(\bar{M})$ with the space of densities on the contact manifold, and fixing the contact volume form, \eqref{dual-1} determines the dual of \eqref{iso} as
\begin{equation}\label{Psi-*}
\Psi^*:  \mathfrak{X}_{\mathrm{con-ham}}^*(\bar{M}) \longrightarrow \mathcal{F}^*(\bar{M}),\qquad \bar{\Pi} \mapsto \bar{f}=\mathrm{div}(\sharp\bar{\Pi}) 
  -   \mathcal{L}_\mathcal{R} \langle \sharp\bar{\Pi},\eta  \rangle     
  + \langle \sharp\bar{\Pi},\eta  \rangle.
\end{equation}
In terms of the Darboux coordinates $(q^i,p_i,z)$, we can compute the density function $\bar{f}$ from \eqref{Psi-*} as follows. Consider a one-form section 
\begin{equation}
\bar{\Pi}  =\bar{\Pi}_i dq^i + \bar{\Pi}^i dp_i + \bar{\Pi}_z dz.
\end{equation}
Let us present the three terms on the right-hand side of \eqref{Psi-*} one by one. 
A direct calculation reads the contact divergence of $\bar{\Pi} $ as 
\begin{equation}
 \mathrm{div}(\sharp\bar{\Pi})  =\frac{\partial \bar{\Pi}^i}{\partial q^i} - \frac{\partial \bar{\Pi}_i}{\partial p_i} -n\bar{\Pi}_z -p_i\frac{\partial \bar{\Pi}_z}{\partial p_i} + \frac{\partial \bar{\Pi}_z}{\partial z} +p_i\frac{\partial \bar{\Pi}^i}{\partial z}.
\end{equation}
Then the second Lie derivative term is computed to be 
\begin{equation}
  \mathcal{L}_\mathcal{R} \langle \sharp\bar{\Pi},\eta  \rangle    = \mathcal{L}_\mathcal{R} \bar{\Pi}_z  =\iota_{R}d\bar{\Pi}_z=\frac{\partial \bar{\Pi}_z}{\partial z}.
\end{equation}
The third term is simply $ \langle \sharp\bar{\Pi},\eta  \rangle = \bar{\Pi}_z$. Adding all of these terms we arrive at the definition of the density function
\begin{equation}
    \bar{f}=  \frac{\partial \bar{\Pi}^i}{\partial q^i} - \frac{\partial \bar{\Pi}_i}{\partial p_i}  -p_i\left(\frac{\partial \bar{\Pi}_z}{\partial p_i}-\frac{\partial \bar{\Pi}^i}{\partial z} \right) -(n-1)\bar{\Pi}_z.
    \end{equation}

\textbf{Coadjoint Flow on $\mathfrak{X}_{\mathrm{con-ham}}^*(\bar{M}).$} Let, as above, $\mathfrak{X}_{\mathrm{con-ham}} (\bar{M})$ be the Lie algebra of contact vector fields with the opposite Jacobi-Lie bracket. That is,
\begin{equation}
ad_{\xi_{\bar{H}}} \xi_{\bar{F}}= - [\xi_{\bar{H}},\xi_{\bar{F}}], 
\end{equation}
which we consider to be the left adjoint action of $\mathfrak{X}_{\mathrm{con-ham}} (\bar{M})$ on itself.  Now dualizing the adjoint action, we arrive at the coadjoint action of $\mathfrak{X}_{\mathrm{con-ham}}(\bar{M})$ on its dual $\mathfrak{X}_{\mathrm{con-ham}}^*(\bar{M})$ as
\begin{equation}
ad^*:\mathfrak{X}_{\mathrm{con-ham}}(\bar{M})\times \mathfrak{X}_{\mathrm{con-ham}}^* (\bar{M})\longrightarrow\mathfrak{X}_{\mathrm{con-ham}}^*(\bar{M}) ,\qquad \langle ad^{\ast }_{\xi_{\bar{H}}} \bar{\Pi}, \xi_{\bar{F}} \rangle = 
 \langle \bar{\Pi} , ad_{\xi_{\bar{H}}} \xi_{\bar{F}} \rangle.
\end{equation}
More explicitly, given an arbitrary field $\xi_{\bar{F}}$ we have 
\begin{equation}
\begin{split}
\langle ad_{\xi_{\bar{H}}}^\ast \bar{\Pi}, \xi_{\bar{F}} \rangle &= \langle \bar{\Pi} , ad_{\xi_{\bar{H}}} \xi_{\bar{F}} \rangle = 
-\int \langle \bar{\Pi},[\xi_{\bar{H}},\xi_{\bar{F}}] \rangle d \mu   \\ &=- \int \langle \bar{\Pi},\mathcal{L}_{\xi_{\bar{H}}} {\xi_{\bar{F}}} \rangle d \mu
 = \int \big\langle \mathcal{L}_{\xi_{\bar{H}}} \bar{\Pi} + \mathrm{div}(\xi_{\bar{H}})\bar{\Pi}, \xi_{\bar{F}}  \big\rangle d \mu 
 \\ &= \int \big\langle \mathcal{L}_{\xi_{\bar{H}}} \bar{\Pi} -  (n+1)  \mathcal{R}(\bar{H})\bar{\Pi}, \xi_{\bar{F}}  \big\rangle d \mu,
\end{split}
\end{equation} 
where we used \eqref{div-X-H} for the divergence of the contact vector field $\xi_{\bar{H}}$. As a result, the coadjoint action may be presented as
\begin{equation}
ad_{\xi_{\bar{H}}}^\ast \bar{\Pi}= \mathcal{L}_{\xi_{\bar{H}}} \bar{\Pi} -  (n+1)  \mathcal{R}(\bar{H})\bar{\Pi}.
\end{equation} 

Being the dual space of a Lie algebra, $\mathfrak{X}_{\mathrm{con-ham}}^*(\bar{M})$ admits a Poisson bracket called the Lie-Poisson bracket \cite{holm11,holm2009geometric,Marsden1999}. More precisely, given two functionals $A$ and $B$ on $\mathfrak{X}_{\mathrm{con-ham}}^*(\bar{M})$ the Lie-Poisson bracket on $\mathfrak{X}_{\mathrm{con-ham}}^*(\bar{M})$ is defined to be 
\begin{equation}\label{LPBr}
\left\{ A,B\right\} ^{\mathfrak{X}_{\mathrm{con-ham}}^*} (\bar{\Pi}
 ) = \int  \Big\langle \bar{\Pi} , ad_{\delta A / \delta \bar{\Pi}} 
 \frac{\delta B}{\delta \bar{\Pi}} \Big\rangle \overline{d \mu} = - \int \Big\langle \bar{\Pi} , \big[\frac{\delta A}{\delta \bar{\Pi}} ,
 \frac{\delta B}{\delta \bar{\Pi}} \big ] \Big\rangle \overline{d \mu}
\end{equation}%
where  ${\delta A}/{\delta \bar{\Pi}}$ stands for the Fr\'echet derivative of the functional $A$.  Given a Hamiltonian functional $\mathcal{H}$, the Lie-Poisson dynamics is governed by the Lie-Poisson equations computed in terms of the coadjoint action, that is,
\begin{equation}\label{LPEq-}
\dot{\bar{\Pi}}= \{\bar{\Pi}, \mathcal{H}\}^{\mathfrak{X}_{\mathrm{con-ham}}^*}=- ad_{\delta \mathcal{H} / \delta \bar{\Pi} }^{\ast }\bar{\Pi}.
\end{equation}
In particular, for the Hamiltonian functional defined by means of the contact vector field $\xi_{\bar{H}}$ as 
 \begin{equation}
 \mathcal{H}(\bar{\Pi})=\int \langle \bar{\Pi},\xi_{\bar{H}} \rangle \overline{d \mu},
\end{equation}
the Fr\'{e}chet derivative $\delta {\bar{H}}/ \delta \bar{\Pi}$ of $\mathcal{H}$ with respect to the momenta becomes the vector field $\xi_{\bar{H}}$. In this case, the Lie-Poisson equation \eqref{LPEq-} takes the form of
\begin{equation} \label{LP-cont-mom}
\dot{\bar{\Pi}} =  - \mathcal{L}_{\xi_{\bar{H}}} \bar{\Pi} +  (n+1)  \mathcal{R}(\bar{H})\bar{\Pi}.
\end{equation} 
Notice that this system governs the dynamics of one-form sections, which reside in the dual space of contact Hamiltonian vector fields, defined on the contact manifold. We refer to \eqref{LP-cont-mom} as the contact momentum-Vlasov equation. It represents an intermediate system that paves the way for what we call the kinetic equations of contact particles. Before examining the kinetic equations in terms of density functions, we will first focus on the strict contact case. Later, we will arrive at the kinetic equation of contact particles and relate it to the contact momentum-Vlasov equation \eqref{LP-cont-mom}. 

\textbf{The Dual Space of Strict Contact Hamiltonian Vector Fields.} Now we consider the algebra \eqref{X-st-con-ham} of strict contact Hamiltonian vector fields
$\mathfrak{X}_{\mathrm{st-con-ham}} ( \bar{M} )$. Similar to the calculation \eqref{dual-1} done above, we compute the precise dual of this vector space by means of $L_2$-pairing. Accordingly, we have  
\begin{equation}\label{dual-1-}
\begin{split}
 \langle \bar{\Sigma} \otimes \overline{d \mu}, Y_{\bar{H}}  \rangle _{L_2} = \int \langle \bar{\Sigma}, Y_{\bar{H}} 
 \rangle   \overline{d \mu}  = 
  \int  \bar{H} \Big( \mathrm{div}(\sharp\bar{\Sigma})     + \langle \sharp\bar{\Sigma},\eta  \rangle \Big)  \overline{d \mu}. 
\end{split}
\end{equation}
Once again, we fix the volume form. Then the non-degeneracy of the pairing \eqref{dual-1-} leads us to define the dual space as
\begin{equation} 
\mathfrak{X}_{\mathrm{st-con-ham}}^* (\bar{M}) = \big \{\bar{\Sigma}  \in 
\Lambda^1 (\bar{M}) : \mathrm{div}(\sharp\bar{\Sigma})   
  + \langle \sharp\bar{\Sigma},\eta  \rangle\neq 0 \big
\}  \cup \{0\}.
\end{equation}
For the contact manifold $\bar{M}$, consider the contactization bundle $\tau:\bar{M}\mapsto M$ over the symplectic base manifold $M$. A real-valued function $H$ on the base manifold can be pulled back to the contact manifold by means of the projection $\tau$. This gives a real-valued function $\tau^*H$ which satisfies $\mathcal{R}(\tau^*H)=0$. 
So that $\tau^*H$ generates a strict contact Hamiltonian vector field $Y_{\tau^*H}$. As a matter of fact, this picture is generic for all functions on the contact manifold that do not depend on the fiber variable. So we arrive at the following isomorphism 
\begin{equation} \label{iso-}
\Gamma: (\mathcal{F}(M)) ,\left\{ \bullet,\bullet \right\} )\longrightarrow 
\left( \mathfrak{X}_{\mathrm{st-con-ham}} ( \bar{M} ) ,-\left[\bullet ,\bullet
\right] \right) , \qquad H \mapsto Y_{\tau^*H}.
\end{equation}
The computation \eqref{dual-1-} provides the dual of this as 
\begin{equation} \label{Gamma-*}
\Gamma^*:\mathfrak{X}^*_{\mathrm{st-con-ham}} ( \bar{M} ) \longrightarrow \mathrm{Den}(M),\qquad \bar{\Sigma}\mapsto   \int _{S^1} \big (\mathrm{div}(\sharp\bar{\Sigma})   
  + \langle \sharp\bar{\Sigma},\eta  \rangle \big ) dz \otimes d \mu,
\end{equation}
where $\mathrm{Den}(M)$ is the space of densities on the symplectic manifold $(M,\Omega)$. Accordingly, we have that the density function
\begin{equation} 
f(q,p)= \int _{S^1} \big (\mathrm{div}(\sharp\bar{\Sigma})   
  + \langle \sharp\bar{\Sigma},\eta  \rangle \big ) dz 
\end{equation}
defined on the base manifold (that is the symplectic manifold) $M$. 
Referring to the Darboux coordinates $(q^i,p_i)$ on $M$, and the induced Darboux coordinates $(q^i,p_i,z)$ on $\bar{M}$ we compute the density function as 
 \begin{equation}
 f(q,p)= \int _{S^1} \Big( \frac{\partial \bar{\Sigma}^i}{\partial q^i} - \frac{\partial \bar{\Sigma}_i}{\partial p_i}  -p_i\left(\frac{\partial \bar{\Sigma}_z}{\partial p_i}-\frac{\partial \bar{\Sigma}^i}{\partial z} \right) + \frac{\partial \bar{\Sigma}_z}{\partial z}-(n-1)\bar{\Sigma}_z \Big) dz 
 \end{equation}
 Note that this distribution function is not in the form of a divergence of a vector field, and thus is not normalized to zero.

\textbf{Coadjoint Flow on $\mathfrak{X}_{\mathrm{st-con-ham}}^*(\bar{M}).$} Let, as above, $\mathfrak{X}_{\mathrm{st-con-ham}} (\bar{M})$ be the Lie algebra of contact vector fields with the opposite Jacobi-Lie bracket. That is,
\begin{equation}
ad_{Y_{\bar{H}}} Y_{\bar{F}}= - [Y_{\bar{H}},Y_{\bar{F}}], 
\end{equation}
which we consider to be the left adjoint action of $\mathfrak{X}_{\mathrm{st-con-ham}} (\bar{M})$ on itself.  Now dualizing the adjoint action, we arrive at the coadjoint action of $\mathfrak{X}_{\mathrm{st-con-ham}}(\bar{M})$ on its dual $\mathfrak{X}_{\mathrm{st-con-ham}}^*(\bar{M})$ as
\begin{equation}
ad^*:\mathfrak{X}_{\mathrm{st-con-ham}}(\bar{M})\times \mathfrak{X}_{\mathrm{st-con-ham}}^* (\bar{M})\longrightarrow \mathfrak{X}_{\mathrm{st-con-ham}}^*(\bar{M}) ,\qquad \langle ad^{\ast }_{\xi_{\bar{H}}} \bar{\Sigma}, \xi_{\bar{F}} \rangle = 
 \langle \bar{\Sigma} , ad_{\xi_{\bar{H}}} \xi_{\bar{F}} \rangle.
\end{equation}
More explicitly, given an arbitrary field $Y_{\bar{F}}$ we have 
\begin{equation} 
\langle ad_{Y_{\bar{H}}}^\ast \bar{\Sigma}, Y_{\bar{F}} \rangle = \langle \bar{\Sigma} , ad_{\xi_{\bar{H}}} \xi_{\bar{F}} \rangle = 
-\int \langle \bar{\Sigma},[Y_{\bar{H}},Y_{\bar{F}}] \rangle \overline{d \mu}   = \int \big\langle \mathcal{L}_{Y_{\bar{H}}} \bar{\Sigma}, Y_{\bar{F}}  \big\rangle \overline{d \mu} .
\end{equation} 
So we arrive at the coadjoint action as
\begin{equation}
ad_{Y_{\bar{H}}}^\ast \bar{\Sigma}= \mathcal{L}_{Y_{\bar{H}}} \bar{\Sigma}.
\end{equation} 
The dual space $\mathfrak{X}_{\mathrm{st-con-ham}}^*(\bar{M})$ has the Lie-Poisson bracket 
\begin{equation}\label{LPBr-st}
\left\{ A,B\right\} ^{\mathfrak{X}_{\mathrm{st-con-ham}}^*} (\bar{\Sigma}
 )  = - \int \Big\langle \bar{\Sigma} , \big[\frac{\delta A}{\delta \bar{\Sigma}} ,
 \frac{\delta B}{\delta \bar{\Sigma}} \big ] \Big\rangle \overline{d \mu}
\end{equation}
and for a given Hamiltonian functional $\mathcal{H}$, the Lie-Poisson dynamics is  
\begin{equation}\label{LPEq-st}
\dot{\bar{\Sigma}}= \{\bar{\Sigma}, \mathcal{H}\}^{\mathfrak{X}_{\mathrm{st-con-ham}}^*}=- ad_{\delta \mathcal{H} / \delta \bar{\Sigma} }^{\ast }\bar{\Sigma}=-\mathcal{L}_{Y_{\bar{H}}} \bar{\Sigma},
\end{equation}
where we chose $ 
 \mathcal{H}(\bar{\Sigma})=\int \langle \bar{\Sigma},Y_{\bar{H}} \rangle \overline{d \mu}$.  
This system governs the dynamics of one-form sections, which are elements of the dual space $\mathfrak{X}_{\mathrm{st-con-ham}}^*(\bar{M})$. We refer to \eqref{LPEq-st} as the strict contact momentum-Vlasov equation. Notice that if the contact Hamiltonian vector field in the contact momentum-Vlasov equation \eqref{LP-cont-mom} does not depend on the action variable $z$, then \eqref{LP-cont-mom} reduces to the strict contact momentum-Vlasov equation \eqref{LPEq-st}. In the following subsection, we will compute the kinetic equations for both contact and strict contact cases in terms of the density variable.

  \subsection{Kinetic Dynamics in Terms of Density Function}

Given a Hamiltonian function $\bar{H}$ on the contact manifold $\bar{M}$, one may define two particle motions on the manifold. One is due to the contact bracket given by $\dot{a}=\{a,\bar{H}\}$, and the other is due to the contact vector field $\xi_{\bar{H}}$ given by $\dot{a}=\xi_{\bar{H}}(a)$. In the symplectic framework, these two definitions coincide but not for the contact geometry. So we treat these two situations one by one. Let us start with the kinetic lift of the contact bracket motion. 

\textbf{Kinetic Lift of Contact Bracket Dynamics.} In view of the contact bracket \eqref{Lag-Bra} of smooth functions, let now  
  \begin{equation}
  ad_{\bar{H}} \bar{K} =\{\bar{H},\bar{K}\}^{(C)}
  \end{equation}
be the adjoint action of $\mathcal{F}(\bar{M})$ on itself. As was noted above, we shall make use of the identification $\mathcal{F}^*(\bar{M})\simeq \mathcal{F}(\bar{M})$ with the dual space. This way, the coadjoint action $\mathcal{F}(\bar{M})$ on $\mathcal{F}^*(\bar{M})$ is computed from
\begin{equation}\label{cont-bra-comt}
\begin{split}
\int \{\bar{F},\bar{H}\}^{(C)}\bar{K} \overline{d \mu} &= \int  \big( \xi_{\bar{H}}(\bar{F})+\bar{F} \mathcal{R}(\bar{H}) \big)  \bar{K}\overline{d \mu} 
\\ &=\int  \bar{K} \big(\iota_{\xi_{\bar{H}}} d\bar{F} \big)      \overline{d \mu} + \int \bar{K}\bar{F} \mathcal{R}(\bar{H}) \overline{d \mu} 
\\ &=\int  \bar{K} d\bar{F}    \wedge  \iota_{\xi_{\bar{H}}} \overline{d \mu} + \int \bar{K}\bar{F} \mathcal{R}(\bar{H}) \overline{d \mu} 
\\ &= - \int  \bar{F}d\bar{K} \wedge  \iota_{\xi_{\bar{H}}} \overline{d \mu}
- \int  \bar{F}\bar{K}  d\iota_{\xi_{\bar{H}}} \overline{d \mu} 
 + \int \bar{K}\bar{F} \mathcal{R}(\bar{H}) \overline{d \mu} 
 \\ &= 
 - \int  \bar{F}(\iota_{\xi_{\bar{H}}} d\bar{K} ) \overline{d \mu}
- \int  \bar{F}\bar{K}  \mathrm{div}(\xi_{\bar{H}}) \overline{d \mu} 
 + \int \bar{K}\bar{F} \mathcal{R}(\bar{H}) \overline{d \mu} 
 \\ &= 
 - \int  \bar{F}\xi_{\bar{H}}(\bar{K}) \overline{d \mu}
+ \int  \bar{F}\bar{K}  (n+1)  \mathcal{R}(\bar{H}) \overline{d \mu} 
 + \int \bar{K}\bar{F} \mathcal{R}(\bar{H}) \overline{d \mu} 
 \\
  &= 
 - \int  \bar{F}\xi_{\bar{H}}(\bar{K}) \overline{d \mu}
+ \int  \bar{F}\bar{K}  (n+1)  \mathcal{R}(\bar{H}) \overline{d \mu} 
 + \int \bar{K}\bar{F} \mathcal{R}(\bar{H}) \overline{d \mu} 
  \\
  &= - \int \bar{F} \big(\{\bar{K},\bar{H}\}^{(C)}-\bar{K}\mathcal{R}(\bar{H})\big)   + (n+2) \int  \bar{F}\bar{K}   \mathcal{R}(\bar{H}) \overline{d \mu} 
    \\
  &=  \int \bar{F} \{\bar{H},\bar{K}\}^{(C)} \overline{d \mu}   +  (n+3) \int  \bar{F}\bar{K}  \mathcal{R}(\bar{H}) \overline{d \mu},
\end{split}
\end{equation}
that is,
 \begin{equation}\label{flip-flow-}
\int \{\bar{F},\bar{H}\}^{(C)}\bar{K} \overline{d \mu} = \int \bar{F} \{\bar{H},\bar{K}\}^{(C)} \overline{d \mu}   +  (n+3) \int  \bar{F}\bar{K}  \mathcal{R}(\bar{H}) \overline{d \mu} 
\end{equation}
for all smooth functions $\bar{F}$, $\bar{H}$, and $\bar{K}$ defined on the contact manifold $\bar{M}$. Accordingly, the coadjoint action appears as 
 \begin{equation} \label{coad-f}
 ad^*_{\bar{H}} \bar{f}=   \{\bar{H},\bar{f}\}^{(C)} - (n+3) \bar{f} \mathcal{R}(\bar{H}).
 \end{equation}
As discussed in the previous subsection, the dynamics on the density level is determined through the coadjoint action. In particular, for the Hamiltonian functional 
   \begin{equation}
\mathcal{H}(\bar{f})=\int \bar{H} \bar{f} d \mu
    \end{equation}
on $\mathcal{F}^*\simeq \mathcal{F}$, where $\bar{H}$ is the Hamiltonian function defined on the extended cotangent bundle, the Fr\'{e}chet derivative $\delta \mathcal{H} / \delta \bar{f} $ becomes $\bar{H}$. In this case, the coadjoint flow may be computed to be
   \begin{equation}
\dot{\bar{f}} = - ad^*_{\delta \mathcal{H} / \delta \bar{f}} \bar{f} =-  ad^*_{\bar{H}} f.
\end{equation}
Substituting the action in \eqref{coad-f} into the coadjoint dynamics, we compute the kinetic equation of contact particles as 
  \begin{equation}\label{LP-con-den}
\dot{\bar{f}}+\{\bar{H},\bar{f}\}^{(C)} =(n+3) \bar{f} \mathcal{R}(\bar{H}).
 \end{equation}
Keeping in mind that the Lie-Poisson dynamics \eqref{LP-cont-mom} in momentum variables and the Lie-Poisson dynamics \eqref{LP-con-den} are related with the Poisson mapping $\bar{\Pi}\mapsto \bar{f}$ given in \eqref{Psi-*}, the kinetic equation is computed in Darboux coordinates as
  \begin{equation}\label{LP-con-den-exp}
  \begin{split}
\frac{\partial \bar{f}}{\partial t} &= -\frac{\partial \bar{H}}{\partial p_i} \frac{\partial \bar{f}}{\partial q^i} + \frac{\partial \bar{H}}{\partial q^i} \frac{\partial \bar{f}}{\partial p_i}
+p_i \left(\frac{\partial \bar{f}}{\partial p_i} \frac{\partial \bar{H}}{\partial z}- \frac{\partial \bar{H}}{\partial p_i} \frac{\partial \bar{f}}{\partial z}\right)\nonumber\\
&\quad+(n+2)\bar{f} \frac{\partial \bar{H}}{\partial z} + \frac{\partial \bar{f}}{\partial z} \bar{H}.
\end{split}
  \end{equation}

 \textbf{Kinetic Lift of Contact V-Field Dynamics.} 
Given a contact vector field $\xi_{\bar{H}}$, let us consider the linear mapping  
 \begin{equation}\label{Psi-mapping}
\Psi_{\xi_{\bar{H}}}:\mathcal{F}(M) \longrightarrow \mathcal{F}(M),\qquad (\xi_{\bar{H}},\bar{K})\mapsto \xi_{\bar{H}}(\bar{K}) 
 \end{equation}
that takes a function to its directional derivative along $\xi_{\bar{H}}$. A similar calculation to the one presented in \eqref{cont-bra-comt} hence yields 
 \begin{equation}\label{flip-flow--}
\int \xi_{\bar{H}}(\bar{F})\bar{K} \overline{d \mu} = \int \bar{F} \{\bar{H},\bar{K}\}^{(C)} \overline{d \mu}   +  (n+2) \int  \bar{F}\bar{K}  \mathcal{R}(\bar{H}) \overline{d \mu} .
\end{equation}
Accordingly, the dual of \eqref{Psi-mapping} is given by 
 \begin{equation}
\Psi^*_{\xi_{\bar{H}}}: \mathcal{F}^*(M) \longrightarrow \mathcal{F}^*(M),\qquad \bar{f}\mapsto \Psi^*_{\xi_{\bar{H}}}(\bar{f}) = \{\bar{H},\bar{f}\}^{(C)} - (n+2) \bar{f}.
 \end{equation}
We then define the dynamics generated by the dual action as
    \begin{equation}\label{LP-con-den-VF}
\dot{\bar{f}} = - \Psi^*_{\xi_{\bar{H}}}(\bar{f}) = - \{\bar{H},\bar{f}\}^{(C)} + (n+2) \bar{f} \mathcal{R}(\bar{H}). 
\end{equation} 
In terms of the Darboux coordinates $(q^i,p_i,z)$, the kinetic dynamics turns out to be
\begin{equation}\label{LP-cont-Psi}
\begin{split}
    \frac{\partial \bar{f}}{\partial t} &= -\frac{\partial \bar{H}}{\partial p_i} \frac{\partial \bar{f}}{\partial q^i} + \frac{\partial \bar{H}}{\partial q^i} \frac{\partial \bar{f}}{\partial p_i}
+p_i \left(\frac{\partial \bar{f}}{\partial p_i} \frac{\partial \bar{H}}{\partial z}- \frac{\partial \bar{H}}{\partial p_i} \frac{\partial \bar{f}}{\partial z}\right) \\
&\quad+(n+1)\bar{f} \frac{\partial \bar{H}}{\partial z} + \frac{\partial \bar{f}}{\partial z} \bar{H}.
\end{split}
\end{equation} 

\textbf{A Direct Calculation to Kinetic Dynamics.}
Instead of geometric constructions, we may use a simplified method of derivation. Evolution of an observable function $a=a(q^i,p_i,z)$ along the vector field  reads
\begin{equation}
\frac{da}{dt} = \xi_{\bar{H}}(a) 
= 
\frac{\partial \bar{H}}{\partial p_i}\frac{\partial a}{\partial q^i}
 -\left(\frac{\partial \bar{H}}{\partial q^i} + p_i \frac{\partial \bar{H}}{\partial z}\right)\frac{\partial a}{\partial p_i}
+\left(-\bar{H} + p_i \frac{\partial \bar{H}}{\partial p_i}\right)\frac{\partial a}{\partial z}.
\end{equation}
In order to construct the kinetic theory, we need to introduce the distribution function $\bar{f}=\bar{f}(q^i,p_i,z)$, which makes it possible to define the averaged functional
\begin{equation}
\bar{A}(\bar{f}) = \int  a(q^i,p_i,z)\bar{f}(q^i,p_i,z) \overline{d \mu}.
\end{equation}
Evolution of this functional is on the one hand given by
\begin{equation}
\frac{d\bar{A}}{dt} = \int  \frac{da}{dt}\bar{f}(q^i,p_i,z) \overline{d \mu},
\end{equation}
while on the other hand, it can also be seen as an evolution of the distribution function itself, 
\begin{equation}
\frac{d\bar{A}}{dt} = \int  a \frac{\partial \bar{f}(q^i,p_i,z)}{\partial t} \overline{d \mu}.
\end{equation}
Rewriting the former expression in the form of the latter (integrating by parts while dropping the boundary terms), we obtain \eqref{LP-cont-Psi}. 

\textbf{Dynamics of Densities for Strict Contact Dynamics.} Let us recall the Poisson mapping in \eqref{Gamma-*}. This turns the contact kinetic dynamics in \eqref{LP-con-den} and in \eqref{LP-con-den-VF} to the Vlasov equation 
\begin{equation}
\dot{f} + \{\bar{H},\bar{f}\}^{(S)} =0 
\end{equation}
where the bracket is the canonical bracket. The dynamics on the Lie algebra dual of quantomorphisms can be thus seen as the standard dynamics of the distribution function on the phase space of particles.

\section{Conclusion: A Hierarchy from Contact to Conformal Dynamics}\label{Sec:Hie}

We have so far provided the generalizations of the Vlasov dynamics for conformal 
 and contact settings on a pure geometrical setting. To sum up and relate the dynamical equations we obtained, we shall present in the present section the hierarchy of the relevant Lie algebras  (by means of Lie algebra homomorphisms) of both the function spaces and the vector fields. We shall then dualize the Lie algebra homomorphisms to arrive at the momentum and Poisson mappings between different levels of descriptions, namely the reversible Hamiltonian dynamics, the conformal Hamiltonian dynamics, and the contact Hamiltonian dynamics. For the level of particle dynamics the relationship between the conformal and the contact Hamiltonian dynamics discussed in, for example, \cite{ghosh2023,Guha18}.

\textbf{Lie Algebra Hierarchy.}
In order to begin with the contact geometry let us first consider the extended cotangent bundle $T^*M\times \mathbb{R}$ (as a contact manifold), along with a contact Hamiltonian  function
\begin{equation}\label{H-barH}
    \bar{H}(q^i,p_i,z)=H(q^i,p_i)-cz
\end{equation}
on it. Then, the contact Hamiltonian dynamics \eqref{conham} takes the particular form 
\begin{equation}\label{contt}
    \frac{dq^{i}}{dt} = \frac{\partial {H}}{\partial p_{i}}, \qquad \frac{dp_{i}}{dt}  = -\frac{\partial H}{\partial q^{i}}+cp_i, \qquad \frac{dz}{dt}=p_i\frac{\partial {H}}{\partial p_{i}}-H(q^i,p_i)+cz.
\end{equation}
The first two equations of \eqref{contt} can be projected to the cotangent bundle $T^*M$, which gives a reduction to the conformal Hamiltonian dynamics \eqref{conformal-Ham-eq}.  
In order to conduct further analysis, we consider two functions  
\begin{equation} 
    \bar{F}(q^i,p_i,z)=F(q^i,p_i)-c_{F}z, \qquad \bar{H}(q^i,p_i,z)=H(q^i,p_i)-c_{H}z,
\end{equation}
and compute their contact bracket \eqref{Lag-Bra} as
\begin{equation}\label{contact-bracket-FbarHbar}
\begin{split}
\{ \bar{F},  \bar{H}\}^{(C)} &= \{ F-c_{F}z, H-c_{H}z\}^{(C)} \\ &= \{ F, H\}^{(C)} -c_{H}\{ F, z\}^{(C)}-c_{F}\{ z, H\}^{(C)}+c_{F}c_{H}\{ z, z\}^{(C)} \\
&=\{ F, H\}^{(S)}-c_{H}\{ F, z\}^{(C)}-c_{F}\{ z, H\}^{(C)}\\
&=\{ F, H\}^{(S)}-c_{H}(F+Z(F))+c_{F}(H+Z(H)),
\end{split}
\end{equation}
where the contact bracket reduces to (the pullback of) the canonical Poisson bracket in the third line.  A direct comparison of \eqref{contact-bracket-FbarHbar} with \eqref{alg-ext} reveals that they look very similar. Accordingly, the choice of the Hamiltonian function \eqref{H-barH} motivates us to determine the Lie algebra homomorphism (more precisely, an embedding) 
\begin{equation}\label{Xi}
  \Xi:  \mathcal{F}(T^*M) \times \mathbb{R} \longrightarrow \mathcal{F}(T^*M\times \mathbb{R}),\qquad (H,c_H) \mapsto \bar{H}(q^i,p_i,z)=H(q^i,p_i)-c_Hz,
\end{equation}
endowing $\mathcal{F}(T^*M) \times \mathbb{R}$ with the Lie algebra bracket in \eqref{alg-ext}, and $ \mathcal{F}(T^*M\times \mathbb{R})$ with the contact bracket in \eqref{Lag-Bra}. 

It is possible to carry this Lie algebra homomorphism to the level of vector fields. To this end, we employ the isomorphisms \eqref{iso} and \eqref{Phi-c} on the domain and the range of \eqref{Xi} in a respected manner to arrive at the Lie algebra homomorphism
 \begin{equation}\label{Upsilon}
\Upsilon: \mathfrak{X}_{\rm{ham}}(T^*M)\times \mathbb{R} \longrightarrow \mathfrak{X}_{\rm{con-ham}}(T^*M\times \mathbb{R}), \qquad X_H^c\mapsto \xi_{\bar{H}}
  \end{equation}
where $\bar{H}$ is the contact Hamiltonian function in \eqref{H-barH}. Finally, in view of the canonical inclusions of the Lie algebras $ \mathfrak{X}_{\rm{ham}}(T^*M)$ and $\mathcal{F}(T^*M) $ into their extensions $ \mathfrak{X}_{\rm{ham}}(T^*M)\times \mathbb{R}$ and $\mathcal{F}(T^*M)\times \mathbb{R} $, we present the following commutative diagram.
\begin{equation}\label{diagram-1}
  \begin{tikzcd}
    \mathfrak{X}_{\rm{ham}}(T^*M)  \arrow[rr,hook] & &\mathfrak{X}_{\rm{ham}}(T^*M)\times \mathbb{R} \arrow[rr,hook,"\Upsilon \text{ in }\eqref{Upsilon}"]   & & \mathfrak{X}_{\rm{con-ham}}(T^*M\times \mathbb{R}) \\ \\
     \mathcal{F}(T^*M) \arrow[uu,"\Phi \text{ in }\eqref{Phi}"]\arrow[rr,hook] & &  \mathcal{F}(T^*M)  \times \mathbb{R} \arrow[rr,hook,swap,"\Xi \text{ in }\eqref{Xi}"]  \arrow[uu,"\Phi^c \text{ in }\eqref{Phi-c}"] & &\mathcal{F}(T^*M\times \mathbb{R})  \arrow[uu,swap,"\Psi \text{ in }\eqref{iso}"]
  \end{tikzcd}
  \end{equation}

\textbf{Poisson Maps Hierarchy.}
Let us next dualize the Lie algebra homomorphisms that appear in the above diagram. To this end, we consider a density function $\bar{f}=\bar{f}(q^i,p_i,z)$ in the dual space $ \mathcal{F}^*(T^*M\times \mathbb{R})$ and examine the mapping $\Xi$ in \eqref{Xi}. We thus obtain the dual mapping
\begin{equation}\label{Xi-dual}
\begin{split}
   \Xi^* &:  \mathcal{F}^*(T^*M\times \mathbb{R}) \longrightarrow \mathcal{F}^*(T^*M) \times \mathbb{R}^* ,\\ &\qquad \qquad 
 \bar{f}(q^i,p_i,z) \mapsto \Big(\int_\mathbb{R} \bar{f}(q^i,p_i,z) dz, \int_{T^*M\times \mathbb{R}} z \bar{f}(q^i,p_i,z) \overline{d \mu}\Big) .  
\end{split}
 \end{equation}
Let us remark that the first term on the range is indeed in $\mathcal{F}^*(T^*M) $, while the second one is a real number in $\mathbb{R}^*\simeq\mathbb{R}$. More precisely, 
 \begin{equation}\label{barf-f}
f(q^i,p_i):=\int_\mathbb{R} \bar{f}(q^i,p_i,z) dz, \qquad c^*:= \int_{T^*M\times \mathbb{R}} z \bar{f}(q^i,p_i,z) \overline{d \mu}.
 \end{equation}
Let us note also that \eqref{Xi-dual} being a dual of a Lie algebra homomorphism, the moments \eqref{barf-f} constitute a Poisson map. Therefore, we can argue that the moments in \eqref{barf-f} map the coadjoint flow \eqref{LP-con-den} on the contact level to the coadjoint flow \eqref{conf-Vlasov-dens} on the conformal Hamiltonian geometry. In terms of one-forms, given $\bar{\Pi}=\bar{\Pi}_i dq^i+\bar{\Pi}^idp_i + \bar{\Pi}_z dz$ we have the projection
\begin{equation}
\Pi _{i} ( q^i,p_i ) =\int_\mathbb{R} \bar{\Pi}_i ( q^i,p_i,z )
dz,\qquad \Pi^i( q^i,p_i ) =\int _\mathbb{R} \bar{\Pi}^i( q^i,p_i,z ) dz,\qquad \bar{\Pi}_z=0.  \label{momo}
\end{equation}%
These maps take the kinetic dynamics in \eqref{LP-cont-mom} to the kinetic dynamics in \eqref{MV-conf}. All these dynamics and projections may now be summarized through the following commutative diagram. 

\begin{equation}
 \xymatrix{
{\begin{array}{c} M-Vlasov \\ \eqref{MV} \end{array}}
\ar[d]_{ \Phi^* ~in ~ \eqref{Phi-*}}
&&{\begin{array}{c} Conformal \\ M-Vlasov \\ \eqref{MV-conf} \end{array}} \ar[ll]^{c=0}\ar[d]^{(\Phi^c)^*~in~\eqref{Phi-c-*}} &&
{\begin{array}{c}Contact \\ M-Vlasov \\ \eqref{LP-cont-mom} \end{array}}
\ar[d]^{\Psi^*~in~\eqref{Psi-*}}\ar[ll]^{\eqref{momo}}
\\
{\begin{array}{c}
Vlasov \\ \eqref{Vlasov-class} \end{array}}  && {\begin{array}{c}Conformal \\ Vlasov \\ \eqref{conf-Vlasov-dens}\end{array}} \ar[ll]^{c=0}
&&
{\begin{array}{c}Contact \\ Vlasov \\ \eqref{LP-cont-Psi}\end{array}}\ar[ll]^{\eqref{barf-f}}
}
\end{equation}

In the future, we would like to apply conformal and kinetic theories in relativistic mechanics and to geometrize non-equilibrium statistical mechanics \cite{JSP2020}.

Another point of view that can be examined in this picture is the study of conformal Hamiltonian systems, where the conformal factor is not constant (as it is in this work), but instead a function of the state variables. On a symplectic manifold $(M,\Omega)$, the dynamics is given by
\begin{equation}
    L_{X_H^c}\Omega = c_H(q^i,p_i)\Omega. 
\end{equation}
After a direct computation, it is immediately apparent that this type of vector field forms a Lie algebra. Hence, the geometric procedure (determining the dual space and writing the Lie-Poisson dynamics) that we employed in this work can be applied to this type of system as well.
Notice also that this type of variable conformal parameter dynamics is implicitly present in the particular case displayed in \eqref{H-barH}, where we relate contact and conformal Hamiltonian dynamics. By simply taking the constant $c$ as a function of the $(q^i,p_i)$ variables, one arrives at this variable conformal Hamiltonian dynamics. We intend to address this in a separate paper, as we wish to analyze the variable conformal Hamiltonian motion in state space within the setting described in \cite{mclachlan2001conformal} such as symmetry and reduction. Then we plan to analyze the Lie algebra (within the framework of Lie algebra extensions), followed by dualization, and hence the kinetic dynamics of the system.
We greatly thank the anonymous referee for highlighting the discussion on the variable conformal parameter and for the critical comments, which have undoubtedly made the paper more accurate and readable.

    \section*{Acknowledgments}
MP was supported by Czech Science Foundation, project 23-05736S.

\appendix
\section{Appendix}\label{Sec:Basics}
\renewcommand{\theequation}{A.\arabic{equation}}

\subsection{Lie-Poisson Dynamics and Coadjoint Flow}\label{App-1}

This very first section of the appendix contains a brief summary of Lie-Poisson dynamics on both the finite dimensional and infinite dimensional Lie groups (to be more precise, diffeomorphism groups).

\textbf{Lie-Poisson Formulation on Lie Groups.}
Let $G$ be a Lie group, for the details and applications to physics of which we refer the reader to \cite{GF,vara84}. Now $e\in G$ denoting the identity element of $G$, the tangent space at the identity element of the group $G$ is called the Lie algebra $\mathfrak{g}:=T_eG$ of $G$. The Lie algebra structure on $\mathfrak{g}$ is determined by a skew-symmetric bilinear bracket $[\bullet,\bullet]$, satisfying the Jacobi identity. 

The representation 
\begin{equation}
ad:\mathfrak{g}\times \mathfrak{g} \mapsto \mathfrak{g},\qquad ad_\xi\eta:=[\xi,\eta]
\end{equation}
of the Lie algebra $\mathfrak{g}$ on itself via its own bracket is called the adjoint representation (action) of the Lie algebra $\mathfrak{g}$ on itself. 

We shall denote by $\mathfrak{g}^*$ the dual space of the Lie algebra $\mathfrak{g}$. Dualizing the adjoint action, then, one arrives at  the coadjoint action of $\mathfrak{g}$ on its dual $\mathfrak{g}^*$, which is given by
\begin{equation}
ad^*:\mathfrak{g}\times \mathfrak{g}^*\longrightarrow \mathfrak{g}^*,\qquad \langle ad^{\ast }_\xi \rho, \eta \rangle = 
 \langle \rho , ad_\xi \eta \rangle.
\end{equation}

The dual space $\mathfrak{g}^*$ admits a Poisson bracket, called the Lie-Poisson bracket, which is defined to be 
\begin{equation}
\left\{ A,B\right\}  (
\rho  ) = \Big\langle \rho ,\left[ \frac{\delta A}{%
\delta \rho},\frac{\delta B}{\delta \rho }\right] \Big\rangle,  \label{LPBr-}
\end{equation}%
see for instance \cite{holm11,holm2009geometric,Marsden1999}, where $\rho$ is an element in the dual space $\mathfrak{g}^*$, $A$ and $B$ are two functionals on $\mathfrak{g}^*$, and the pairing on the right-hand side is the natural pairing between  $\mathfrak{g}^*$ and $\mathfrak{g}$. Notice also that ${\delta A}/{\delta \rho }$ stands for the Fr\'echet derivative of the functional $A$. Assuming the reflexivity on vector spaces, we view ${\delta A}/{\delta \rho }$ in $\mathfrak{g}$, justifying thus the Lie bracket that appears on the right-hand side of  \eqref{LPBr-}. Let us note further that the right-hand side of \eqref{LPBr-} with the opposite sign would still define a Poisson algebra. We shall, however, prefer the above convention, and justify our choice in the following paragraph. 

Given a Hamiltonian functional $H$, the dynamics is governed by the Lie-Poisson equations computed in terms of the coadjoint action as
\begin{equation}
\dot{\rho}= - ad_{\delta H / \delta \rho }^{\ast }\rho. 
\label{LPEq}
\end{equation}
Let $\mathfrak{g}$ and $ \mathfrak{h}$ be two Lie algebras and let $\phi:\mathfrak{g}\mapsto \mathfrak{h}$ be a Lie algebra homomorphism, that is,
\begin{equation}
\phi[\xi,\eta]=[\phi(\xi),\phi(\eta)],
\end{equation}
for any $\xi$ and $\eta$ in $\mathfrak{g}$. 

The dual spaces $\mathfrak{g}^*$ and $\mathfrak{h}^*$ are Lie-Poisson spaces and the dual mapping $\phi^*: \mathfrak{h}^*\mapsto \mathfrak{g}^*$ is a momentum and a Poisson mapping \cite{Marsden1999}. The relation between the coadjoint representations is computed to be 
\begin{equation}\label{coad-coad}
\phi^*\circ ad^*_{\phi(\xi)} =ad^*_{\xi}\circ \phi^* 
\end{equation}
for all $\xi$ in $\mathfrak{g}$.

\textbf{Lie-Poisson Dynamics for Diffeomorphism Group.}
For many continuous and kinetic theories including fluid flows and plasma theories, configuration spaces are diffeomorphism groups which are infinite-dimensional Lie groups \cite{ArKh98,ebin70,marsden82}. To see this, we start with a bunch of particles resting in a (volume) manifold $M$. We denote the set of all diffeomorphisms on $M$ by $\mathrm{Diff}(M)$ \cite{banyaga97}. The motion of the particles is determined by the left action of $\mathrm{Diff}(M)$ on the particle space $M$. The right action commutes with the particle motion and constitutes
an infinite-dimensional symmetry group 
called the particle relabelling symmetry. The Lie algebra of $\mathrm{Diff}(M)$ is the space  of vector fields $\mathfrak{X}(M)$, where the Lie algebra bracket is the opposite Jacobi-Lie bracket of vector fields, that is, 
\begin{equation} \label{Lie-brr}
ad_X Y = [X,Y]_{\mathfrak{X}\left( M%
\right)}=-[X,Y]_{JL}=-\mathcal{L}_X Y,
\end{equation}
with $\mathcal{L}_X$ being the Lie derivative operator. 
We define the dual space $\mathfrak{X}^{\ast }(M)$ of the Lie algebra as the space of one-form densities $\Lambda ^{1}(M) \otimes \mathrm{Den}(M)$ on $M$, where the pairing between a vector field $X$ and a dual element $\Pi\otimes d \mu$ is defined to be the $L_2$-pairing (simply multiply-and-integrate form)
\begin{equation}\label{L-2}
\langle \bullet, \bullet \rangle_{L_2}: \Lambda^1(M)\otimes \mathrm{Den}(M)\times \mathfrak{X}(M)\longrightarrow \mathbb{R}, \qquad (\Pi\otimes d \mu,X)\mapsto \int_{M} \langle\Pi , X \rangle   d \mu.   
\end{equation}
The pairing inside the integral is the one between the one-form $\Pi$ and the vector field $X$, and $d \mu$ is a density (a volume form) on $M$.

To compute the coadjoint action of the Lie algebra onto the dual space, we perform the following calculation
\begin{equation}
\begin{split}
\langle ad_X^\ast (\Pi\otimes d \mu), Y \rangle &= \langle \Pi\otimes d \mu, ad_X Y\rangle = - \int_M \langle \Pi,\mathcal{L}_XY \rangle d \mu
\\
&= \int_M \big\langle \mathcal{L}_X \Pi + \mathrm{div}(X)\Pi, Y \big\rangle d \mu 
\end{split}
\end{equation} 
where $\mathrm{div}(X)$ stands for the divergence of the vector field with respect to the volume form $d \mu$. To write the second line of this calculation, we have used integration by parts. Hence,
\begin{equation}\label{coad-gen}
ad_{X}^{\ast }\left( \Pi \otimes d \mu\right) =\big( \mathcal{L}_{X}\Pi
+\mathrm{div}(X) \Pi \big) \otimes d \mu,
\end{equation}%
where $\mathrm{div}(X)$
denotes the divergence of the vector field $X$ with respect to the volume
form $d \mu$. At this point, without loss of generalization, we fix the volume form $d \mu$, so that we particularly consider a dual element as a one-form $\Pi$. 

Now we consider a particle that moves according to the dynamics generated by a vector field $X$ defined on the manifold $M$. This particle motion can be lifted to the evolution of distribution functions as follows. Consider a linear Hamiltonian functional on the space of one-form densities $\Lambda^1(M)\otimes \mathrm{Den}(M)$ given by
 \begin{equation}
 H(\Pi \otimes d \mu)=\int_M \langle \Pi,X \rangle d \mu,
\end{equation}
where $d \mu$ is a volume form. 
Then $\delta H/ \delta \Pi$ being the Fr\'{e}chet derivative of $H$ with respect to the momenta, the Lie-Poisson equation turns out to be
\begin{equation}
\dot{\Pi} \otimes d \mu=-ad^*_{\delta H/\delta \Pi} (\Pi \otimes d \mu) = -ad^*_{X} (\Pi \otimes d \mu). 
\end{equation}
Now we fix the volume $d \mu$ and recall the coadjoint action given in \eqref{coad-gen}, which gives the Lie-Poisson equation
\begin{equation}\label{LP-gen}
\dot{\Pi}=-\mathcal{L}_{X}\Pi - \mathrm{div}(X) \Pi.   
\end{equation}
If the dynamics is generated by a divergence-free vector field (for example, the case of incompressible fluid flow, or Vlasov flow), then the second term on the right-hand side of
\eqref{LP-gen} drops, and we obtain
\begin{equation}
\dot{\Pi}=-\mathcal{L}_{X}\Pi.  \label{LP}
\end{equation}

\subsection{Double Cross Sum Lie Algebras} \label{AnAlgebraExtension}

The present section of the appendix contains the construction of a Lie algebra out of two Lie algebras $\mathfrak{g}$ and $\mathfrak{h}$, admitting mutual actions 
\begin{equation}
    \triangleright : \mathfrak{h}\otimes \mathfrak{g} \to  \mathfrak{g}, \quad \eta \otimes \xi \mapsto \eta \triangleright \xi, \qquad \triangleleft : \mathfrak{h} \otimes \mathfrak{g} \to \mathfrak{h}, \quad \eta \otimes \xi=\eta \triangleleft \xi,
\end{equation}
which are assumed to satisfy
\begin{equation}
\begin{split}
    \eta \vartriangleright \lbrack \xi _{1},\xi _{2}]=[\eta \vartriangleright
\xi _{1},\xi _{2}]+[\xi _{1},\eta \vartriangleright \xi _{2}]+(\eta
\vartriangleleft \xi _{1})\vartriangleright \xi _{2}-(\eta \vartriangleleft
\xi _{2})\vartriangleright \xi _{1}, \\
\lbrack \eta _{1},\eta _{2}]\vartriangleleft\xi =[\eta _{1},\eta _{2}\vartriangleleft\xi ]+[\eta _{1}%
\vartriangleleft\xi ,\eta _{2}]+\eta _{1}\vartriangleleft (\eta _{2}\vartriangleright \xi
)-\eta _{2}\vartriangleleft (\eta _{1}\vartriangleright \xi ).
\end{split}
\end{equation}
Such a pair $(\mathfrak{g},\mathfrak{h})$ of Lie algebras is called a \emph{matched pair} of Lie algebras, see for instance \cite{Ma90}, see also \cite{thiffeault2000}. Then the vector space direct sum $\mathfrak{g} \oplus \mathfrak{h}$ happens to be a Lie algebra along with the bracket given by 
$ \mathfrak{g} \oplus\mathfrak{h}$ as 
\begin{equation}\label{mathcpairbracket}
  \left[ (\xi_{1}, \eta_{1} ), (\xi_{2},\eta_{2}) \right] = (\left[ \xi_1,\xi_2\right]+\eta_1\triangleright\xi_2-\eta_2\triangleright \xi_1, \left[ \eta_1,\eta_2\right] + \eta_1\triangleleft \xi_2- \eta_2 \triangleleft \xi_1 ),  
\end{equation}
for any $(\xi_1,\eta_1),(\xi_2,\eta_2) \in \mathfrak{g} \oplus \mathfrak{h}$. The Lie algebra $\mathfrak{g} \bowtie \mathfrak{h} := \mathfrak{g} \oplus \mathfrak{h}$ is called the \emph{double cross sum} of the pair $(\mathfrak{g},\mathfrak{h})$.

Double cross sum construction extends the semi-direct sum construction. Indeed, choosing for instance the right action in \eqref{mathcpairbracket} to be trivial, that is $\eta\triangleleft \xi=0$ for all $\xi$ in $\mathfrak{g}$ and $\eta $ in $\mathfrak{h}$, and letting the Lie algebra $\mathfrak{h}$ to be trivial, that is $\left[ \eta_1,\eta_2\right]=0$ for all $\eta_1$ and $ \eta_2$ in  $\mathfrak{h}$, we see at once that the bracket \eqref{mathcpairbracket} reduces to
\begin{equation} \label{Alg}
  \left[ (\xi_{1}, \eta_{1} ), (\xi_{2},\eta_{2}) \right] = (\left[ \xi_1,\xi_2\right]+\eta_1\triangleright\xi_2-\eta_2\triangleright \xi_1, 0).
\end{equation}

Physically the double cross sum construction corresponds to the collective motion of two dynamical systems. The Lagrangian (Euler-Poincar\'{e}) dynamics on double cross sum Lie algebras were studied in \cite{esen2021matched, EsenKudeSutlu21, esen2017lagrangian}, and it is followed by the analysis of the Hamiltonian dynamics on the dual spaces of double cross sum Lie algebras, \cite{esen2016hamiltonian}. The case of discrete dynamics, on the other hand, has been treated in \cite{esen2018matched}. The present work, along these lines, concerns the coadjoint flow on the dual space of a double cross sum Lie algebra.

In order to formulate the coadjoint action of a double cross sum Lie algebra on its dual space, we shall now fix a number of notations. 

Fixing the algebra element $\xi \in \mathfrak{g}$ in the left action we define the linear operation
\begin{equation}
    \mathfrak{b}_{\xi}: \mathfrak{h} \to \mathfrak{g},  \quad  \eta \mapsto \eta \triangleright \xi.
\end{equation}

Then, for an arbitrary $\zeta \in \mathfrak{g}^*$, the dual of the mapping $\mathfrak{b}_\xi$ is obtained to be
\begin{equation}
\mathfrak{b_\xi}^*: \mathfrak{g}^* \to \mathfrak{h}^*, \quad
\langle  \mathfrak{b}^*_{\xi} \zeta, \eta \rangle =\langle \zeta, \mathfrak{b}_{\xi}\eta \rangle.
\end{equation}
Next, we dualize the left action of $\mathfrak{h}$ on $\mathfrak{g}$ into a dual right action of $\mathfrak{h}$ on $\mathfrak{g}^\ast$, which may be given as
\begin{equation}
    \overset{\ast}{\triangleleft}~\eta: \mathfrak{g}^{*} \to \mathfrak{g}^{*}, \qquad \zeta \to \zeta \overset{\ast}{\triangleleft} \eta, \qquad \langle \zeta \overset{\ast}{\triangleleft} \eta, \xi\rangle=\langle \zeta, \eta \triangleright \xi \rangle.
\end{equation}
The linear algebraic dual of the adjoint action gives the coadjoint action. The dual of \eqref{Alg} is then given by
\begin{equation}
\begin{split}
    \langle ad^*_{(\xi_1,\eta_1)}(\zeta,\nu), (\xi_2,\eta_2) \rangle & =   \langle (\zeta,\nu), ad_{(\xi_1,\eta_1)}(\xi_2,\eta_2) \rangle
   \\  &=\langle (\zeta,\nu),  (\left[ \xi_1,\xi_2\right]+\eta_1\triangleright\xi_2-\eta_2\triangleright \xi_1, 0) \rangle 
   \\ &=\langle \zeta,  \left[ \xi_1,\xi_2\right]+\eta_1\triangleright\xi_2-\eta_2\triangleright \xi_1   \rangle 
   \\ &=\langle ad_{\xi_1}^*\zeta , \xi_2 \rangle +  \langle \zeta \overset{\ast}{\triangleleft} \eta_1,  \xi_2 \rangle 
   - \langle \mathfrak{b}^*_{\xi_1}\zeta, \eta_2   \rangle .
       \end{split}
\end{equation}
To sum up, for given any $(\xi,\eta)\in \mathfrak{g}\oplus \mathfrak{h}$ and any $(\zeta,\nu)\in\mathfrak{g}^*\oplus \mathfrak{h}^*$, the coadjoint action may be formulated as
\begin{equation}\label{coadjointaction}
   ad^*_{(\xi,\eta)}(\zeta,\nu)= (ad^*_\xi \zeta+\zeta\overset{\ast}{\triangleleft}\eta, -\mathfrak{b}^*_\xi \zeta).
\end{equation}
  Finally, given a Hamiltonian functional $H=H(\zeta, \nu )$ on the direct sum $\mathfrak{g}^*\oplus \mathfrak{h}^*$, the Lie-Poisson equations \eqref{LPEq} turn out to be 
\begin{equation} \label{Sultan}
    \begin{split}
        &\dot{\zeta}=-ad^*_{\frac{\delta H}{\delta \zeta }}\zeta-\zeta\overset{\ast}{\triangleleft}\frac{\delta H}{\delta \nu },\\
        & \dot{\nu}=\mathfrak{b}^*_\frac{\delta H}{\delta \zeta}\zeta,
    \end{split}
\end{equation}
which is the abstract evolution equation for $\zeta$ and $\nu$.

\subsection{From Momentum to Density Formulations of Conformal Kinetic Theories}\label{link-link}

We link now the conformal kinetic equations \eqref{MV-conf} in momentum formulation and the conformal kinetic equation \eqref{conf-Vlasov-dens} in terms of the density.  To this end, we start with the density function $f$ given in \eqref{f-c-defn} and compute its time derivative in view of the conformal kinetic equations \eqref{MV-conf} in momentum formulation. This reads
  \begin{equation}
     \begin{split}
\frac{\partial f}{\partial t}&=\mathrm{div} \Omega ^\sharp(\dot{\Pi})= - (\mathrm{div} \Omega ^\sharp(\mathcal{L}_{X_H^c}\Pi + c_Hn  \Pi)) \\&=- \mathrm{div} \Omega ^\sharp\big(\mathcal{L}_{X_H^c}\Pi\big)-(c_Hn)\mathrm{div} \Omega ^\sharp 
 (\Pi),
        \end{split}
    \end{equation}  
where, keeping \eqref{f-c-defn} in mind, the second term is equal to $-c_Hnf$ whereas the first one needs a more detailed observation. To write the first term as a function of $f$ we pair it with an arbitrary function using the $L_2$-pairing and compute
      \begin{equation}\label{calc-long}
       \begin{split}
       \int_M   &\mathrm{div} \Omega ^\sharp\big(\mathcal{L}_{X_H^c}\Pi\big) K d \mu 
       =     
       \int_M   \langle \mathcal{L}_{X_H^c}\Pi, X_K \rangle d \mu  
         =     
       \int_M   \langle \mathcal{L}_{X_H-c_HZ}\Pi, X_K \rangle d \mu 
              \\
       &=  \int_M   \langle \mathcal{L}_{X_H}\Pi, X_K \rangle d \mu -  c_H \int_M   \langle \mathcal{L}_{Z}\Pi, X_K \rangle d \mu
       \\
       &=  - \int_M   \langle \Pi, \mathcal{L}_{X_H}X_K \rangle d \mu  +  c_H \int_M   \langle \Pi, \mathcal{L}_{Z}X_K \rangle d \mu + c_H \int_M   \langle \Pi, X_K \rangle \mathrm{div}(Z)d \mu
        \\
       &=
        \int_M   \langle \Pi,  X_{\{H,K \}^{(S)}}\rangle d \mu  +  c_H \int_M   \langle \Pi, X_{Z(K)+K} \rangle d \mu + c_H  \int_M   \langle \Pi, X_K \rangle \mathrm{div}(Z)d \mu
               \\
       &=
        \int_M  \mathrm{div} \Omega ^\sharp(\Pi)\{H,K \} ^{(S)}d \mu  +  c_H \int_M   \langle \Pi, X_{Z(K)+K} \rangle d \mu - c_H n\int_M   \langle \Pi, X_K \rangle d \mu
               \\
       &= \int_M  \{\mathrm{div} \Omega ^\sharp(\Pi),H \} ^{(S)}K d \mu  +  c_H \int_M   \langle \Pi, X_{Z(K)} \rangle d \mu - c_H (n-1) \int_M   \langle \Pi, X_K \rangle d \mu
           \\
       &= \int_M  \{\mathrm{div} \Omega ^\sharp(\Pi),H \}^{(S)} K d \mu  +  c_H \int_M  \mathrm{div}  \Omega ^\sharp(\Pi) Z(K)  d \mu - c_H (n-1)  \int_M   \mathrm{div} \Omega ^\sharp(\Pi) K  d \mu 
          \\
       &=\int_M  \{\mathrm{div} \Omega ^\sharp(\Pi),H \}^{(S)} K d \mu - c_H \int_M  Z(\mathrm{div}  \Omega ^\sharp(\Pi)) K  d \mu - c_H \int_M  (\mathrm{div}  \Omega ^\sharp(\Pi)) K\mathrm{div}(Z)   d \mu
       \\&\hspace{3cm}  - c_H (n-1) \int_M   \mathrm{div} \Omega ^\sharp(\Pi) K  d \mu 
       \\
       &=\int_M  \{f,H \}^{(S)} K d \mu - c_H \int_M  Z(f) K  d \mu + c_H \int_M  f K n   d \mu  - c_H (n-1)  \int_M   f K  d \mu 
           \\
       &=\int_M \big( \{f,H \}^{(S)} - c_H Z(f)+ c_H f \big) K d \mu.
       \end{split}
          \end{equation} 
As a result, we have 
       \begin{equation}
       \mathrm{div} \Omega ^\sharp\big(\mathcal{L}_{X_H^c}\Pi\big)=\{f,H \}^{(S)} - c_H Z(f) + c_H f,
        \end{equation}
via which we obtain
     \begin{equation}
         \begin{split}
\frac{\partial f}{\partial t}&=\mathrm{div} \Omega ^\sharp(\dot{\Pi})= - \mathrm{div} \Omega ^\sharp\big(\mathcal{L}_{X_H^c}\Pi\big)-(c_Hn)\mathrm{div} \Omega ^\sharp 
 (\Pi) 
 \\&= \{H,f \}^{(S)}+c_H Z(f)-c_H (n+1) f.
 \end{split}
    \end{equation}
    This is exactly the same as the evolution of the density variable given in the first line of \eqref{conf-Vlasov-dens}. Let us perform a similar analysis for the real variable as well. We thus compute the time derivative of the scalar variable established in \eqref{f-c-defn} to arrive at
         \begin{equation}  
         \begin{split}
      \frac{\partial c^{*}}{\partial t} &= -\int_{M} \langle \dot{\Pi},Z \rangle d \mu = \int_{M}\langle\mathcal{L}_{X_H^c}\Pi,Z \rangle d \mu + \int_{M}
      \langle cn \Pi, Z \rangle d \mu \\
      &=\int_{M}\langle\mathcal{L}_{X_H-cZ}\Pi,Z \rangle d \mu + \int_{M}
      \langle cn \Pi, Z \rangle d \mu \\
     &= \int_{M}\langle\mathcal{L}_{X_H}\Pi,Z \rangle d \mu-c\int_{M}\langle\mathcal{L}_{Z}\Pi,Z \rangle d \mu + \int_{M}
      \langle cn \Pi, Z \rangle d \mu
      \\
     &=-\int_{M}\langle\Pi,\mathcal{L}_{X_H}Z \rangle d \mu + c\int_{M}\langle\Pi,\mathcal{L}_{Z}Z \rangle d \mu 
     +c\int_{M}\langle\Pi,Z \rangle \mathrm{div}(Z)d \mu + \int_{M}
      \langle cn \Pi, Z \rangle d \mu
       \\
     &=\int_{M}\langle\Pi,X_{Z(H)+H}\rangle d \mu=
     \int_{M}\mathrm{div}\Omega^\sharp(\Pi) (Z(H)+H)d \mu =  \int_{M}f (Z(H)+H)d \mu,
 \end{split}      
    \end{equation}
which coincides with the evolution of the real variable given in the second line of \eqref{conf-Vlasov-dens}. So, the conformal kinetic equation \eqref{conf-Vlasov-dens} becomes a particular instance of the abstract Lie-Poisson equation \eqref{Sultan}.

\bibliographystyle{abbrv}
\bibliography{references}

\end{document}